\documentclass[a4paper,11pt]{article}

\usepackage{a4wide}
\usepackage[latin1]{inputenc}
\usepackage[T1]{fontenc}
\usepackage[english]{babel}
\usepackage{graphicx,subfigure}
\usepackage{geometry}
\usepackage{bm}
\usepackage{color}
\usepackage{latexsym}
\usepackage{amssymb}
\usepackage[numbers]{natbib}
\begin{document}

\begin{center}
\textbf{\Large{Multi-Scale Turbulence Injector: a new tool to generate intense homogeneous
and isotropic turbulence for premixed combustion}}
\end{center}

\begin{center}
\large{Nicolas Mazellier\footnote{Corresponding authors: nicolas.mazellier@univ-orleans.fr and
bruno.renou@coria.fr}\footnote{Permanent
address: Institut PRISME,
8 rue L{\'e}onard de Vinci, 45072 Orl{\'e}ans, France \vspace{6pt}},
Luminita Danaila and Bruno Renou$^\ast$}
\end{center}

\begin{center}
\small{UMR 6614 CORIA, INSA et Universit{\'e} de Rouen, BP 08, 76801 St.
Etienne du Rouvray, France}
\end{center}

\thispagestyle{empty}

\paragraph{Abstract} Nearly homogeneous and isotropic, highly
turbulent flow, generated by an original multi-scale
injector is experimentally studied. This multi-scale injector
is made of three perforated plates shifted in space
such that the diameter of their holes and their blockage ratio
increase with the downstream distance. The Multi-Scale Turbulence Injector
(hereafter, {\it MuSTI}) is compared with a Mono-Scale Turbulence Injector
({\it MoSTI}), the latter being constituted by only the last
plate of {\it MuSTI}. This comparison is done for both cold and
reactive flows.

For the cold flow, it is shown that, in comparison with the
classical mono-scale injector, for the {\it MuSTI} injector:
\textit{(i)} the turbulent kinetic energy is roughly twice larger,
and the kinetic energy supply  is distributed over the whole range
of scales. This is emphasized by second and third order 
structure functions. \textit{(ii)}  the transverse fluxes of momentum and
energy are enhanced, \textit{(iii)} the homogeneity  and isotropy are
reached earlier ($\approx 50$\%), \textit{(iv)} the jet merging
distance is the relevant scaling length-scale of the turbulent flow,
\textit{(v)} high turbulence intensity ($\approx 15$\%) is achieved
in the homogeneous and isotropic region, although the Reynolds
number based on the Taylor microscale remains moderate ($Re_\lambda
\approx 80$).

In a second part, the interaction between the multi-scale generated
turbulence and the premixed flame front is investigated by laser
tomography. A lean V-shaped methane/air flame is stabilised on a
heated rod in the homogeneous and isotropic region of the turbulent
flow. The main observation is that the flame wrinkling is hugely
amplified with the multi-scale generated injector, as testified by
the increase of the flame brush thickness.\bigskip


\section{Introduction}

In the framework of turbulent premixed combustion, the
morphology and the properties of the flame are essential
inputs for combustion models. In particular, finite rate
chemistry effects as well as flame front wrinkling and
straining resulting from the competition between turbulence
and combustion have to be accurately modeled to produce
reliable predictions \citep{Pitsch2006}. These requirements led to
the concept of the so-called combustion diagrams (see e.g.
\citep{Peters1986} and \citep{BorghiChampion2000}) which are derived
from a phenomenological approach comparing the different time-scales
and length-scales involved in the turbulence/combustion
competition. Despite of their doubtless interest in combustion
modeling, combustion diagrams are based on severe assumptions:

\begin{enumerate}
\item[(i)] the turbulence is considered as being non affected by
heat release (frozen turbulence). However, the complex interaction
between turbulent structures and the flame front (pre-heated and
reaction zones) is not well interpreted by standard combustion
diagrams \citep{Poinsotetal1990}.
\item[(ii)] the turbulent flow is considered as homogeneous and
isotropic. This assumption allows to describe the turbulent flow
properties using a reduced set of parameters (usually large-scale
variables).
\item[(iii)] criteria and regime limits are not determined from
accurate estimations but rely on order-of-magnitude approximations.
\end{enumerate}

One of the objectives of works dedicated to turbulent premixed
combustion was to improve the concept of combustion diagrams
and therefore to lead to relevant combustion models. Experimentally,
huge efforts were made to reproduce the different
flame regimes. Unfortunately, this task requires to generate
highly turbulent flows for which the homogeneity and isotropy
assumptions are not fulfilled \citep{OYoungBilger1997, Dinkelackeretal1998,
Dunnetal2007}. Numerically, recent advances in Direct Numerical
Simulation (e.g. \citep{Vervischetal2004, Nadaetal2004}) provide
a much deeper comprehension of the combustion physics by
investigating the local flame structure. The confrontation
of such approaches with reliable experimental data represents
therefore a promising way to bring new insights in combustion modeling.
From the experimental point of view, the main challenge
is to conciliate highly turbulent level with homogeneity and
isotropy properties.
\vspace*{1ex}  

Grid-generated turbulence, commonly recognised as nearly homogeneous
and isotropic, has received a large attention over the past century
since the precursory work of Batchelor \& Townsend
\citep{BatchelorTownsend1948}. Although this kind of flow has been
extensively studied, it remains an ideal (and easy to be produced)
candidate to identify and model the underlying physics of turbulence
and combustion. Corrsin \citep{Corrsin1963a} provided a comprehensive review of
grid-generated turbulence, which is produced by the passage of a
stream through a grid with a specific pattern. The flow contraction
imposed by the section reduction (with a blockage ratio $\sigma$ quantifying
the ratio between the blockage area and the wind tunnel's section) at
the grid location induces mean velocity gradients, implying
turbulent kinetic energy production. Far away from the grid,
turbulence decays and becomes nearly homogeneous and isotropic. By
adding a secondary contraction downstream the grid, Comte-Bellot \&
Corrsin \citep{ComtebellotCorrsin1966} succeeded in improving global
isotropy. The main disadvantage of standard grid-generated
turbulence lies in its inability to reach high turbulence intensity
(few percents) and therefore is restricted to moderate Taylor-based
Reynolds numbers $Re_\lambda$.
\vspace*{1ex}

Numerous attempts have been dedicated to produce homogeneous and
isotropic turbulence at high Reynolds numbers. Gad-el-Hak \& Corrsin
\citep{GadelHakCorrsion1974} developed a grid equipped with jets
injecting air either in co-flowing or in counter-flowing configurations.
They found that the counter-flowing configuration was able to
produce a more intense turbulence (with $Re_\lambda \approx 150$)
than the standard grid, by also keeping a good homogeneity and
isotropy. In the nineties, Makita \citep{Makita1991} and later
Mydlarsky \& Warhaft \citep{MW1996} succeeded in generating highly
turbulent flow using active grids ($Re_\lambda \approx 150 - 1000$).
Active grids are made of small wings mounted on a standard grid frame.
The rotation of the wings is controlled by independent step motors
which can be driven either in synchronous or random mode. Active grids
produce an isotropy level which is slightly worse than that of
the standard grid-generated turbulence. However, broad inertial
ranges were revealed on energy spectra, attesting of the efficiency
of such a device. More recently, Hurst \& Vassilicos \citep{HurstVassilicos2007}
investigated turbulence generated by passive fractal grids made from
the reproduction of a given pattern at different scales. For a
square pattern, they reported high values of turbulence
intensity (8\%), as well as large Reynolds numbers ($Re_\lambda
\approx 150 - 450$) although the blockage ratio of the fractal grids ($\approx$ 25\%)
is much lower than standard regular grids ($\approx$ 34\%). The isotropy level
was comparable to that of active grids and the homogeneous and the
isotropic region appeared further away than for the standard
grid-generated turbulence.
\vspace*{1ex}

The purpose of the present work is to create experimentally a
nearly homogeneous isotropic turbulence,  with a large turbulence
intensity. We investigate an original turbulence generator made of
the combination of several perforated plates. It is worth to mention
that turbulence manipulation by grid/screen
combination is not a new concept. Tan-Atichat \& al.
\citep{TanAtichatetal1982} and later on Groth \& Johansson
\citep{GrothJohansson1988} investigated the turbulence reduction with
combination of screens and perforated plates. The efficiency of their turbulence
\textit{``manipulators''} was tested for various upstream
conditions. They found that the energy decay was accelerated through
the interaction between the incoming turbulence and the small-scale
turbulence generated by the \textit{``manipulators''}. However, to our
knowledge, no attempt has been devoted to amplify
turbulence by using grid/screen combination, which is the aim of the
present work.
\vspace*{1ex}

The paper is organized
as follows. The experimental facility and the measurement techniques
are described in section 2. The characteristics of both the original
multi-scale injector and a reference injector are presented in
section 3. The main properties of turbulence at large- and small-scales,
as well as the interaction with premixed combustion are discussed
in section 4.

%%%%%%%%%%%%%%%%%%%%%%%%%%%%%%%%%%%%%%%%%%%%%%%%%%%%%%%%%%%%%%%%%%%%%%%%%%%%%%%%%%%%%%%
%%%%%%%%%%%%%%%%%%%%%%%%%%%%%%%%%%%%%%%%%%%%%%%%%%%%%%%%%%%%%%%%%%%%%%%%%%%%%%%%%%%%%%%
%%%%%%%%%%%%%%%%%%%%%%%%%%%%%%%%%%%%%%%%%%%%%%%%%%%%%%%%%%%%%%%%%%%%%%%%%%%%%%%%%%%%%%%

\section{The experimental set-up}

\subsection{The wind-tunnel facility}

\begin{figure}[h]
\centering
\includegraphics[width=10cm]{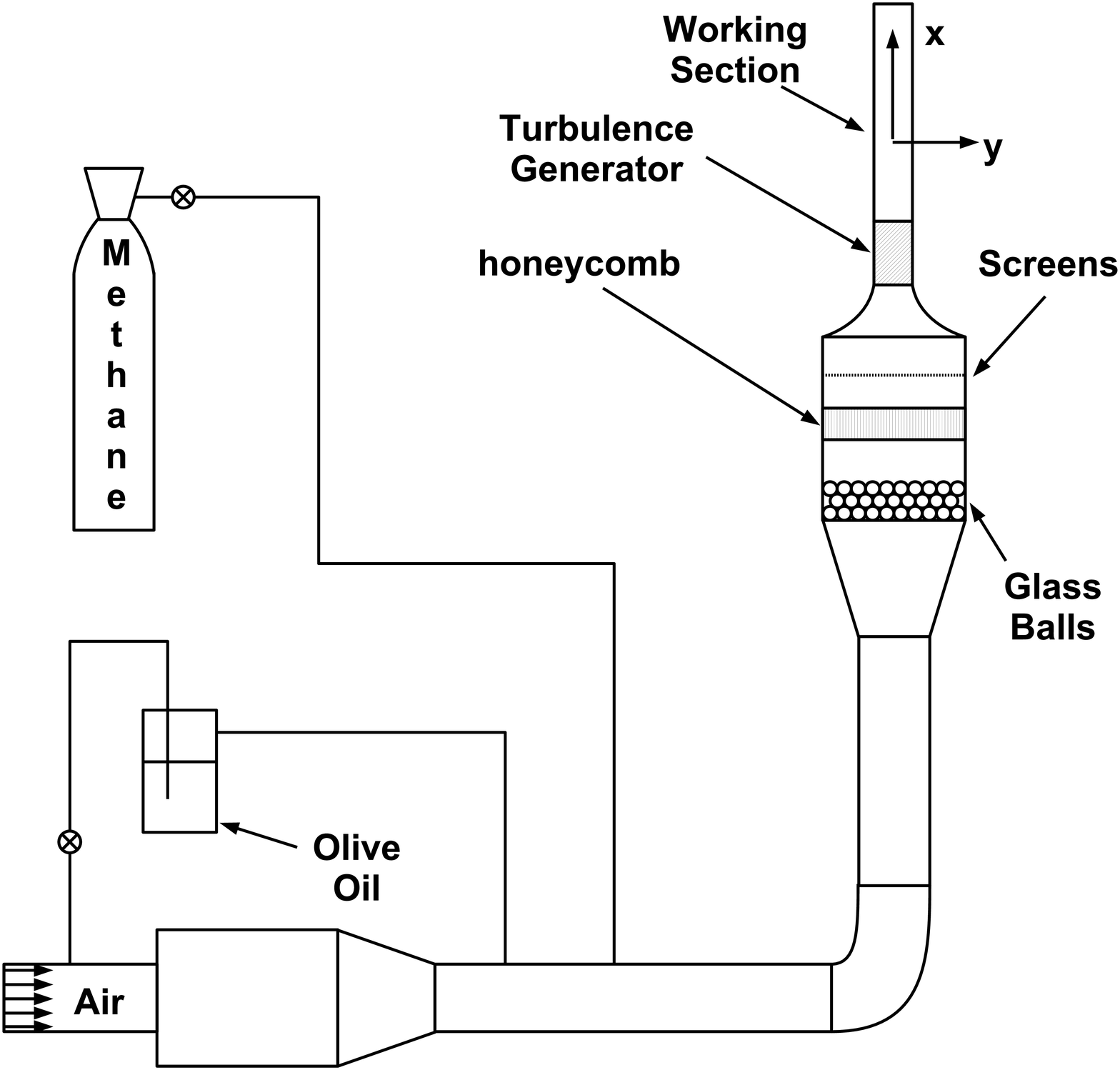}
\caption{\textit{Schematic representation of the wind-tunnel facility.}}
\label{fig:tunnel}
\end{figure}

Experiments were carried out in an open-loop vertical wind-tunnel
with a square test section ($8 \times 8 \mbox{cm}^2$) adapted to the study
of steady combustion \citep{Degardinetal2006}, see Figure
\ref{fig:tunnel}. The test section is $40$cm long and is
equipped with optical access. In non-reactive configuration, the
working fluid is air supplied by a network of compressed air,
regulated via a mass flow meter/controller Bronkhorst (F-206AI).

The flow is directed to a divergent-convergent settling chamber
constituted of glass balls bed, honeycombs and screens to attenuate
residual turbulent perturbations. The flow is then accelerated
through a (11:1) aspect ratio contraction. The turbulence
generators are set between the outlet of the convergent and the
inlet of the working section. In the absence of any obstacle, the
residual turbulence level is as low as $0.4$\% in the working
section. The inlet velocity $U_\infty$ computed from the mean flow
rate and the test section area is imposed to $3.7$m/s.

In the following, we use the notation $(u, v, w)$ for denoting the
velocity components in the $(x, y, z)$ directions. The origin of the
streamwise direction $x$ is taken at the outlet of the turbulence
generator.

\subsection{Velocity measurements}

\begin{figure}[htbp]
\centering \subfigure {\includegraphics[width=5cm]{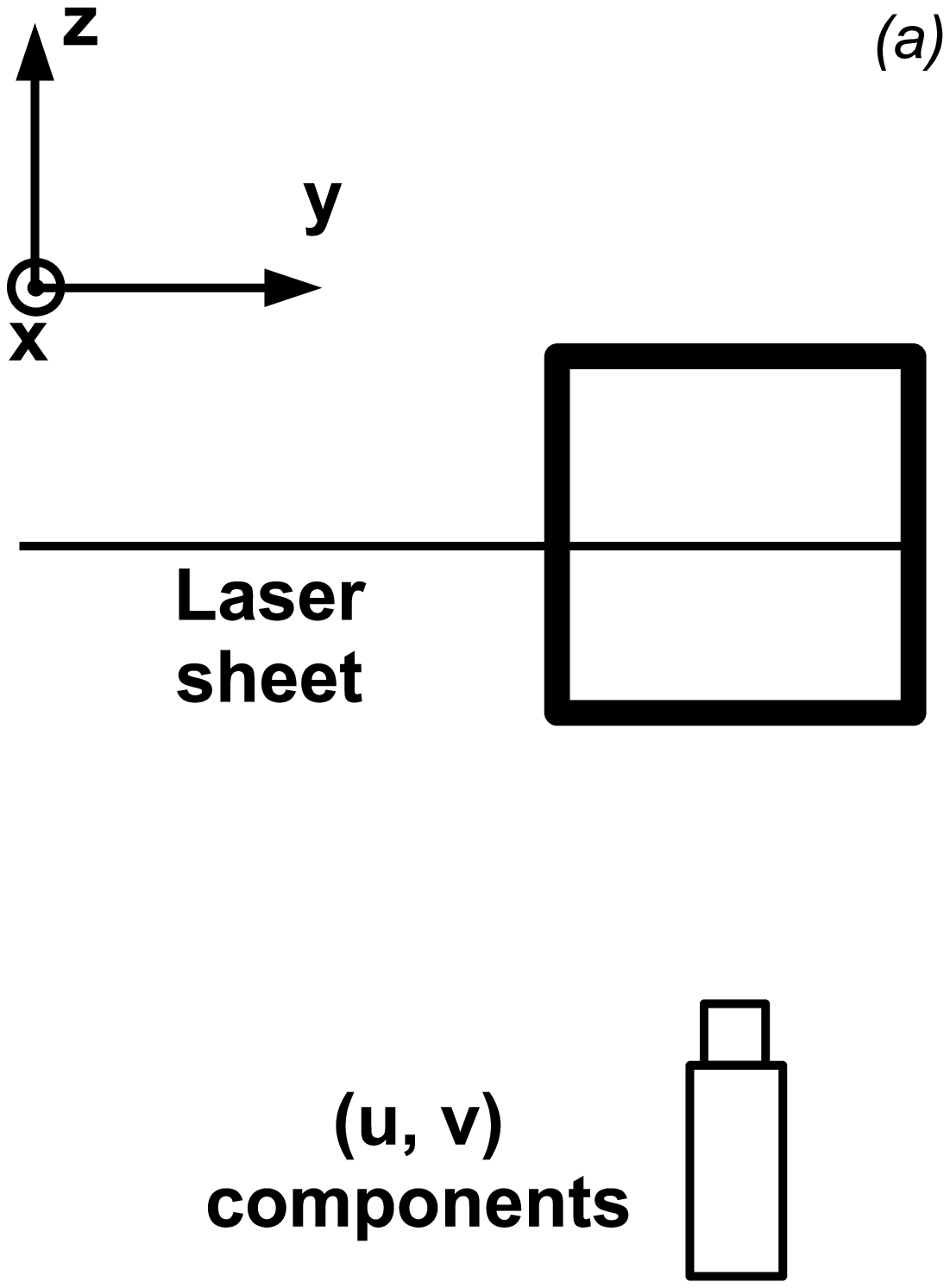}
\label{fig:PIV}} \hspace{2cm} \subfigure
{\includegraphics[width=5cm]{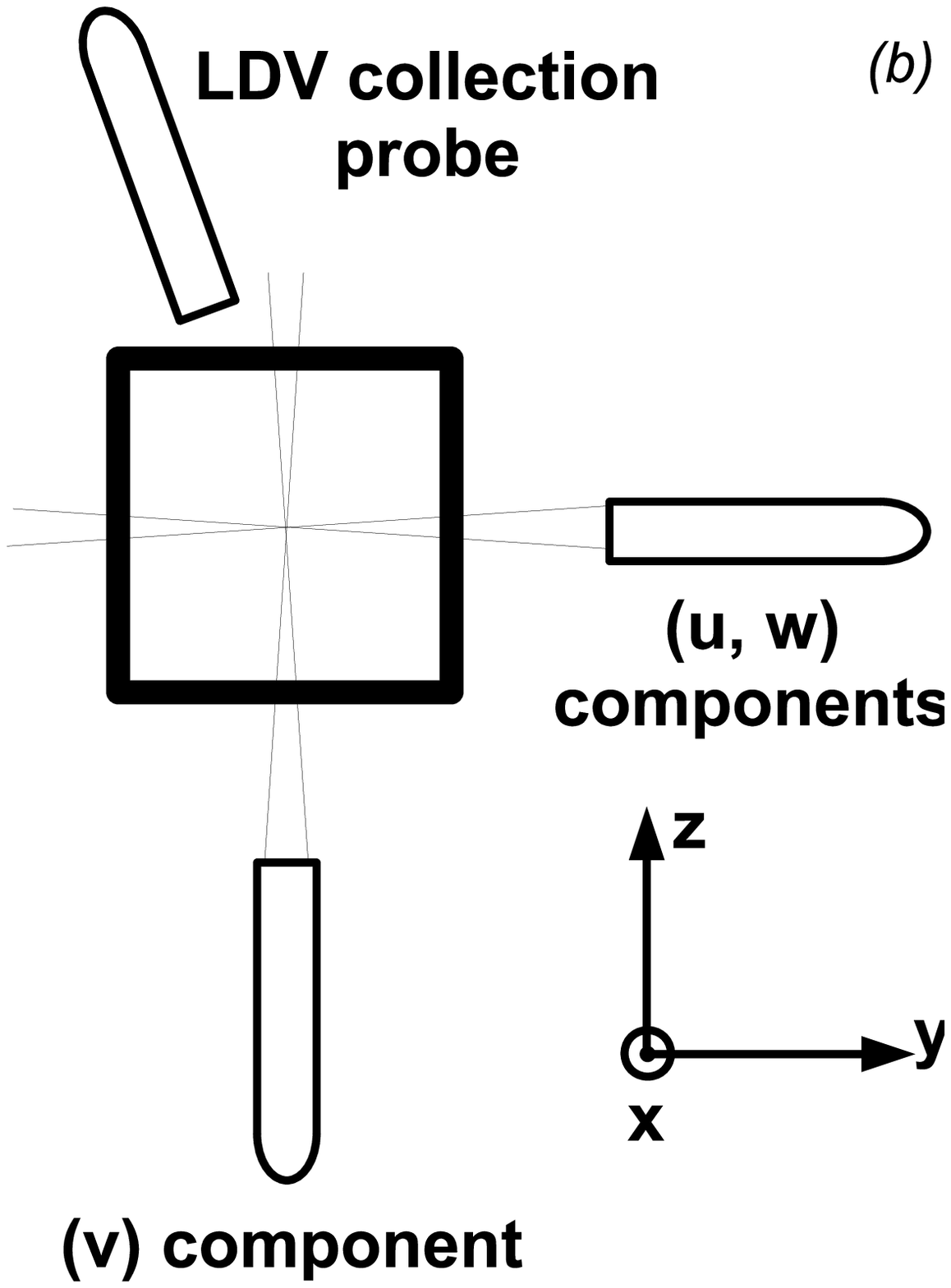} \label{fig:LDV}}
\caption{Top-views of the \textit{PIV} \textit{(a)} and the
\textit{LDV} \textit{(b)} configurations.}
\end{figure}

Velocity field is investigated by two independent optical methods:
Particle Image Velocimetry (\textit{PIV}) and Laser Doppler
Velocimetry (\textit{LDV}). A schematic representation of the
experimental set-up is given in Figures \ref{fig:PIV} and
\ref{fig:LDV}. During measurements, the wind-tunnel is seeded with
olive oil droplets generated by a particle-seeding apparatus and
injected far upstream of the working section.  The average diameter
$d_p$ of the olive oil particles is close to $1\mu$m, as separately
calibrated with a Malvern diffractometer. The Stokes number $S_k =
\tau_p f_\eta$ comparing the particle response time $\tau_{p}=\frac{d_p^2}
{18 \nu}$ (with $\nu$ the kinematic viscosity of the fluid) and the
Kolmogorov's frequency $f_\eta = \frac{U}{2 \pi \eta}$ (with $\eta$ the
Kolmogorov scale and $U$ the mean flow velocity), is much lower than 0.1
ensuring a good flow tracing. The seeding rate was adjusted to ensure
good droplet density and homogeneity during optical measurements.

\subsubsection{PIV system}
The spatial velocity field in non-reactive configuration is investigated
via \textit{PIV} with a Nd-Yag laser (Big Sky laser, $120$ mJ/pulse, $532$ nm)
as light source. The vertical laser sheet coincides with the middle plane
of the working section corresponding to the ($x,y$)-plane (see Figure \ref{fig:PIV}).
Light scattered from olive oil droplets is collected on a CCD camera
(FlowMaster LaVision, 12-bits, $1280 \times 1024 \mbox{pix}^2$) with a 50 mm f/1.2 Nikkor
lens. The magnification ratio is of $16$ pix/mm, which leads to a physical field of 
view of $80 \mbox{mm} \times 64 \mbox{mm}$. Particle images are post-processed with
the standard commercial package available in Davis 6.2 (LaVision
Company). Velocity field computation is based on a multi-pass
algorithm with adaptive window deformation. The starting and final
interrogation windows are imposed to $64 \mbox{pix}^2$ and $16 \mbox{pix}^2$
respectively with $50$\% overlapping giving a final map of $160 \times
128$ vectors. According to \citep{Foucaultetal2004}, the spatial resolution
of the \textit{PIV} system is about $2.25$mm. We
estimate that the final interrogation window ranges between $10$ and
$25$ times bigger than the Kolmogorov scale $\eta$. Due to the poor
spatial resolution, PIV results will be used to investigate the
large scales of the flow. To ensure statistical convergence, a total
of $3500$ images are acquired for each configuration. Both optics
and camera are mounted on a 3D traversing system allowing to
displace the investigated frame along the entire working section. An
overlap of almost $1/3$ between the different frames is imposed in
order to check the continuity of statistics.

\vspace*{0.2cm}

\subsubsection{LDV system}
Three-component \textit{LDV}, with a $8W$ Argon-ion laser ($514.5
$nm, $488 $nm and $476.5 $nm) as light source, is used to measure
local velocity statistics with a good temporal resolution. Data are
collected in forward scatter mode at $25^\circ$ off axis (see Figure
\ref{fig:LDV}) and processed with an IFA755 processor (TSI) set in a
non-coincident single measurement per burst mode. Signals are then
digitized and stored on computer hard-drive via ICA-LDV32.net
software. For each measurement, several parameters (laser intensity,
photomultiplier intensity, filtering window ...) are carefully
adjusted to maximize the mean data rate $f_s$ which ranges between
$12 $kHz and $22 $kHz. The minimum sampling rate $f_s$ varies
between $0.6$ and $2.6$ times the estimated Kolmogorov's frequency
$f_\eta = \frac{U}{2 \pi \eta}$. Each optic fiber is clamped on
independent high precision linear stages allowing to
superimpose accurately the three measurement volumes. The
\textit{LDV} spatial resolution $\frac{\ell}{\eta}$ (with $\ell$ the
typical size of the \textit{LDV} measurement volume) is estimated to
range between $0.6$ and $1.8$, which is at least $10$ times better
than \textit{PIV}. For each station, $8 \times 10^6$ samples are
throughout acquired (almost $1/3$ per channel). The associated
measurement time ranges between $5 \times 10^4$ and $25 \times 10^4$
integral time-scale ensuring the statistical convergence. The
\textit{LDV} system is mounted on a 3D traversing system controlled
by computer. Measurement points are taken along the tunnel
centreline and downstream the centre between two successive holes.

\subsection{Flame structure characterization}

\begin{figure}[htbp]
\centering \subfigure {\includegraphics[width=5cm]{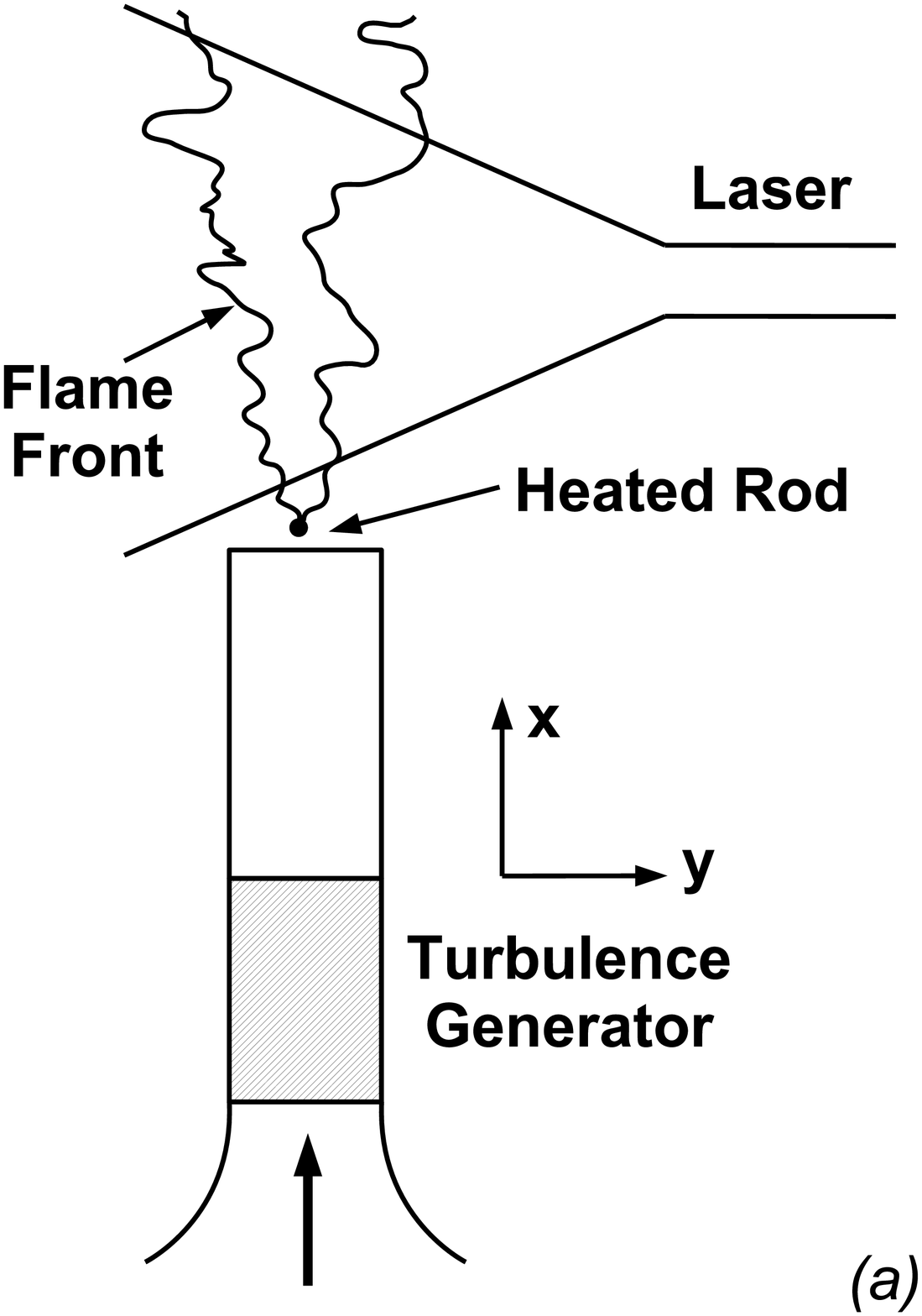}
\label{fig:Tomo}} \hspace{0.5cm}
%\subfigure
%{\includegraphics[width=4cm]{TomoEx.eps} \label{fig:TomoEx}}
%\hspace{0.5cm}
\subfigure
{\includegraphics[width=4cm]{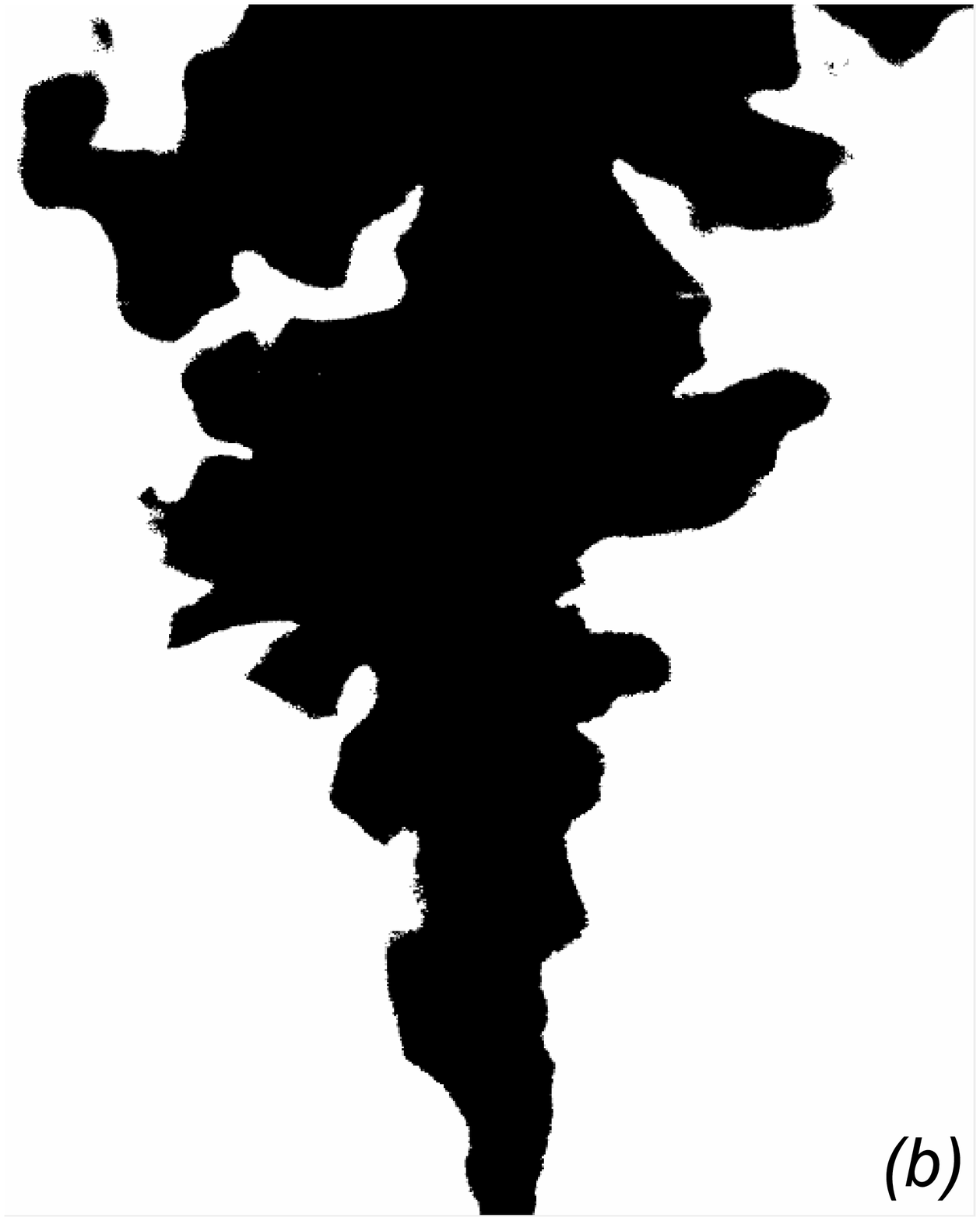} \label{fig:TomoExBin}}
\caption{\textit{(a)} Front-view of the \textit{Tomography}
configuration. \textit{(b)} Typical example of binary flame front
image obtained by thresholding.}
\end{figure}

In reactive configuration, methane/air mixture is used with an
equivalent ratio $\phi = 0.6$. A 2D stationary V-shaped flame is
stabilised on a tiny heated rod ($1 $mm in diameter) placed in the
middle plane of the test section (see Figure \ref{fig:Tomo}). The
heated rod position, denoted $x_{rod}$, is located in the
homogeneous and isotropic region of the turbulent flow. Downstream
the rod location, the lateral walls of the tunnel are removed to
avoid flame/wall interaction. In order to minimize the effect of
lateral mixing layers on flame properties, we focus on the very
near-field of the heated rod. The flames are visualised by laser
sheet tomography with the same light source as for \textit{PIV}. The
reacting flow was seeded with olive oil droplets which evaporate at the
entrance of the flame front. The instantaneous two-dimensional flame
surface was obtained by differentiation, on the flame recordings, of
the dark and bright areas, representing burned and unburned states
respectively. The flame images are recorded with a \textit{PIV}
camera using identical magnification ratio with that of the
\textit{PIV}. For each image, the contours of the turbulent flame
(which physically correspond to the instantaneous location of the
isotherm $500$K) are extracted with an edge detection algorithm
using an adaptive and smoothing procedure. Figure \ref{fig:TomoExBin}
shows a typical example of the resulting binarized image where white areas
correspond to fresh gases, while black regions represent the burnt
gases (and therefore the flame). A total of 1000 images were
acquired to determine the Reynolds average properties of the flame.

%%%%%%%%%%%%%%%%%%%%%%%%%%%%%%%%%%%%%%%%%%%%%%%%%%%%%%%%%%%%%%%%%%%%%%%%%%%%%%%%%%%%%%%%%%%%%%%%%%
%%%%%%%%%%%%%%%%%%%%%%%%%%%%%%%%%%%%%%%%%%%%%%%%%%%%%%%%%%%%%%%%%%%%%%%%%%%%%%%%%%%%%%%%%%%%%%%%%%
%%%%%%%%%%%%%%%%%%%%%%%%%%%%%%%%%%%%%%%%%%%%%%%%%%%%%%%%%%%%%%%%%%%%%%%%%%%%%%%%%%%%%%%%%%%%%%%%%%

\section{The turbulence generators}

\subsection{The reference injector}

\begin{figure}[htbp]
\centering \subfigure {\includegraphics[width=5cm]{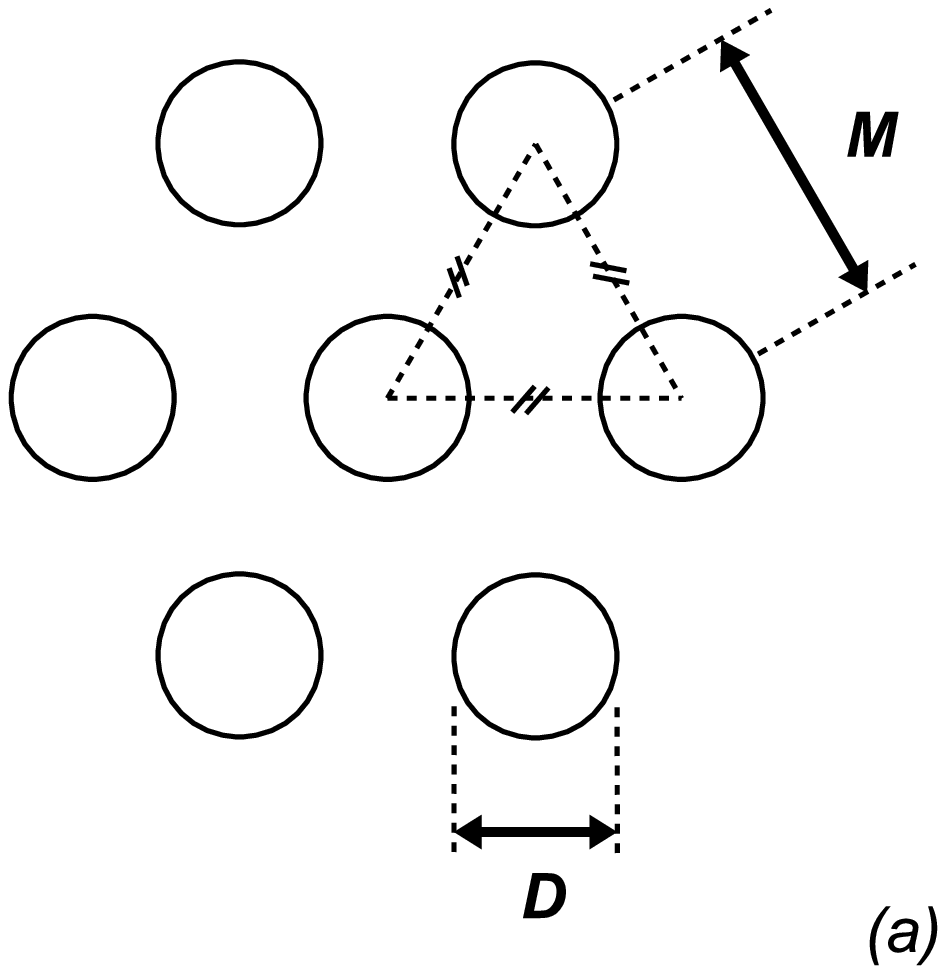}
\label{fig:pattern}} \hspace{2cm} \subfigure
{\includegraphics[width=5cm]{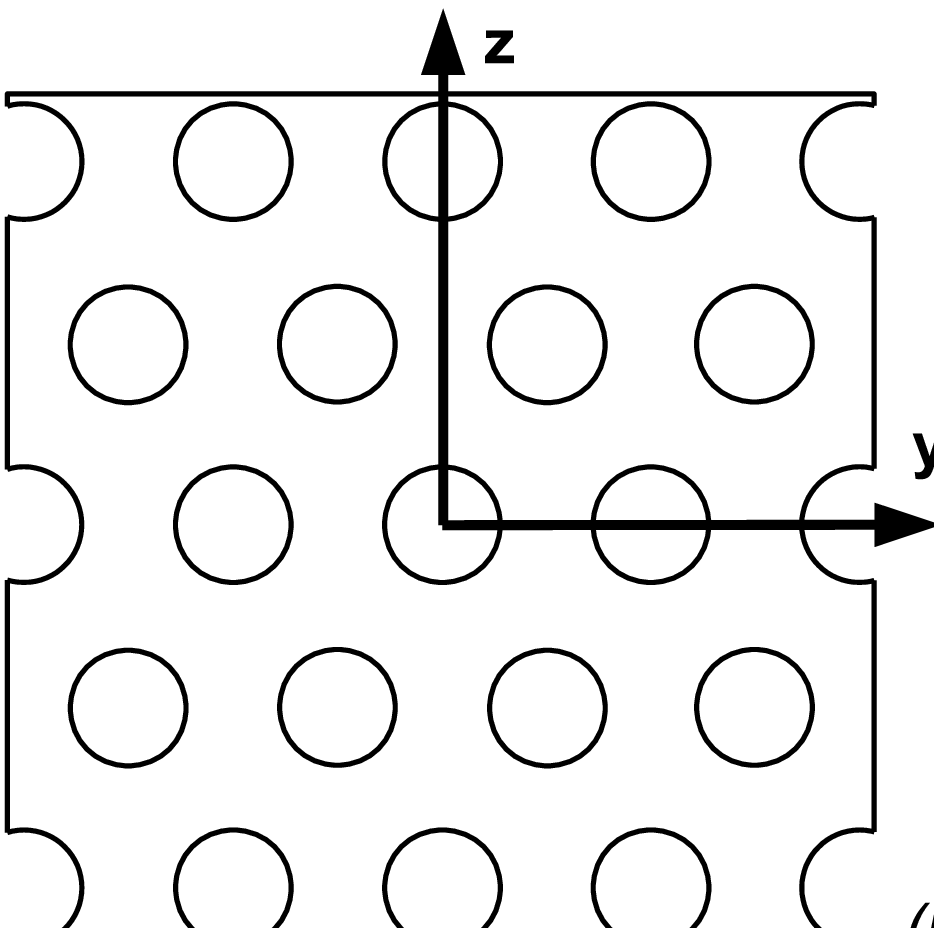} \label{fig:grid}}
\caption{\textit{(a)} Hole pattern. \textit{(b)} Typical example of
perforated plate.}
\end{figure}

In the present work, turbulence is generated by perforated plates either
alone or in combination via a multi-scale injector which is the novelty
of the paper. Each plate is characterised by both a hole diameter $D$
and the mesh size $M$ and spans the entire wind tunnel. The holes are
circular and arranged in a triangular network as shown in Figure
\ref{fig:pattern}. Moreover, the perforation is straight over the entire
thickness of the plates and the holes network is chosen such that the
tunnel's centreline coincide with a hole centre. A typical illustration
of a perforated plate is given in Figure \ref{fig:grid}.

The amount of turbulent kinetic energy injected by each plate is controlled
by the pressure drop $\Delta p$ \citep{LawsLivesey1978}:

\begin{equation}
\Delta p = \frac{1}{2} f(Re) K_\sigma \rho U^2
\label{eq:pressure}
\end{equation}

with $\rho$ the fluid density, $U$ the mean velocity in the $x$ direction,
$f(Re)$ an empirical function and

\begin{equation}
K_\sigma = \frac{1 - (1 - \sigma)^2}{(1 - \sigma)^2}.
\label{eq:Ksigma}
\end{equation}

The function $f(Re)$ becomes constant ($\approx 0.5$ for screens)
for high enough Reynolds number \citep{GrothJohansson1988}.
\vspace*{1ex}

Prior to the manufacturing of the new multi-scale device, a single perforated
plate was chosen and used as a reference injector. The geometrical properties
of this reference injector are given in Table \ref{tab:MoSTI}. By opposition
to the multi-scale injection, the reference injector produces a mono-scale
forcing of the flow and relies on the so-called turbulent energy cascade to
transfer the injected energy from large-scales to small-scales. The reference
injector is therefore called \textit{MoSTI} (Mono-Scale Turbulence Injector)
in the following.

\begin{table}[htbp]
\begin{center}
\begin{tabular}{cccc}
\hline
$D$ (mm) &  $M$ (mm)    &   $\sigma$    & $K_\sigma$ \\
\hline
\hline
$15$    &   $24.7$  &   $0.67$  &   $8.18$ \\
\hline
\end{tabular}
\caption{Geometrical characteristics of the mono-scale injector (MoSTI)}
\label{tab:MoSTI}
\end{center}
\end{table}

Considering technical constraints (tunnel's size for instance), the mesh size $M$
has been fixed to about $1/3$ of the tunnel side. The blockage ratio,
$\sigma = 0.67$, is situated at midway between that usually reported for bi-plane grids
($\sigma \approx 0.34$) \citep{ComtebellotCorrsin1966} and that used by Villermaux \&
Hopfinger ($\sigma \approx 0.9$) \citep{VillermauxHopfinger1994}. The
latter observed sustained jet-oscillations over large distance
downstream the grids that are not present in our study.

%The turbulence generated by the \textit{MoSTI} has been studied (see hereafter) for
%inlet velocity $U_\infty = 3.7m/s$. The integral length-scale $\Lambda_u$, Taylor's
%microscale $\lambda$ and Kolmogorov's microscale $\eta$ have been evaluated in the
%homogeneous and isotropic region of the flow (see section ... for more details). Their
%values are reported in Table \ref{tab:MoSTI}.

\subsection{The multi-scale injection concept}

The multi-scale injector we have designed is built from a
combination of $N$ perforated plates. In the
present work, for simplicity we have chosen only three plates ($N = 3$).
Moreover, one of them is the same as that for the \textit{MoSTI}.

\begin{figure}[htbp]
\centering
\includegraphics[width=10cm]{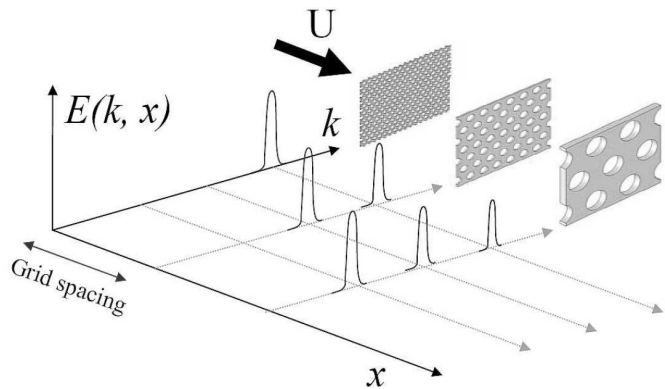}
\caption{Schematic representation of the Multi-Scale injector
(\textit{MuSTI}).} \label{fig:MuSIG}
\end{figure}

As shown in Figure \ref{fig:MuSIG}, the three plates are shifted
in space such that both hole's diameter $D_j$ and mesh size $M_j$ increase
in the mean-flow direction (where subscript refers to the $j^{th}$ plate).
This device generates a space-delayed
(or time-delayed) multi-scale forcing of the flow and is therefore
refered to as \textit{MuSTI} (Multi-Scale Turbulence Injector) in
the following. The concept of this multi-scale injection can be easily
represented in wavenumber space (see Figure \ref{fig:MuSIG}). Following
the flow, small-scales (high wavenumbers) are first excited, then
medium-scales and finally large-scales (low wavenumbers). Using this
arrangement, we aim to experimentally produce a multi-scale
forcing in wavenumber space, thus mimicking a cascade process.
In other words, each perforated plate $j$ is
expected to inject a specific spectral energy $E_j(k)$ at given
injection wavenumbers $k$. This approach has
already been developed, for instance, by Mazzi \& Vassilicos
\citep{MazziVassilicos2004} who performed continuous and discrete
fractal forcing in stationary Direct Numerical Simulations. The
authors revealed that the turbulence generated was very sensitive to
the properties of the forcing. From Figure \ref{fig:MuSIG}, one can
remark that the complete design of the multi-scale injector requires three
independent geometrical parameters:

\begin{enumerate}
\item the mesh size $M_j$,
\item the spacing $L_j$ between plates $j$ and $j+1$,
\item the pressure drop constant $K_{\sigma j}$.
\end{enumerate}

It is important to notice that the hole's diameter $D_j$ and the blockage
ratio $\sigma_j$ are functions of both $M_j$ and $K_{\sigma j}$. 

\subsection{The final design of the multi-scale injector}

The geometrical parameters of the last perforated plate being fixed (identical
to those of the \textit{MoSTI}), it remains to choose those of the two first
plates. The approach we have adopted to choose the triad ($M_j, K_{\sigma j}$
and  $L_j$) of those plates has been based on the investigation of the turbulence
generated downstream by various perforated plates used alone.

\subsubsection{The mesh size $M_j$}

\begin{figure}[htbp]
\centering
\subfigure
{\includegraphics[width=7.5cm]{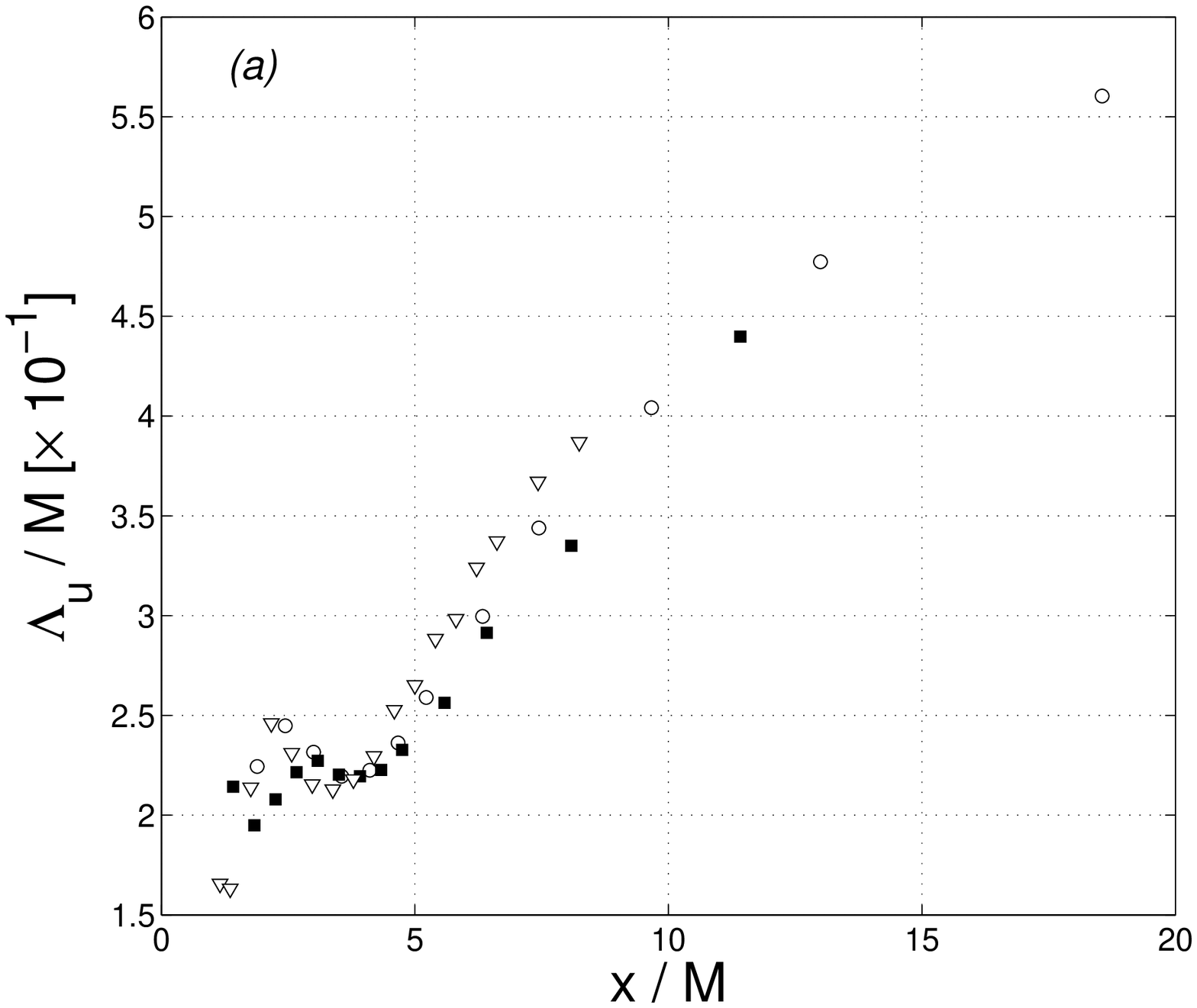}
\label{fig:IntegralScaleMonoNorm}}
%\hspace{0.01cm}
\subfigure
{\includegraphics[width=7.5cm]{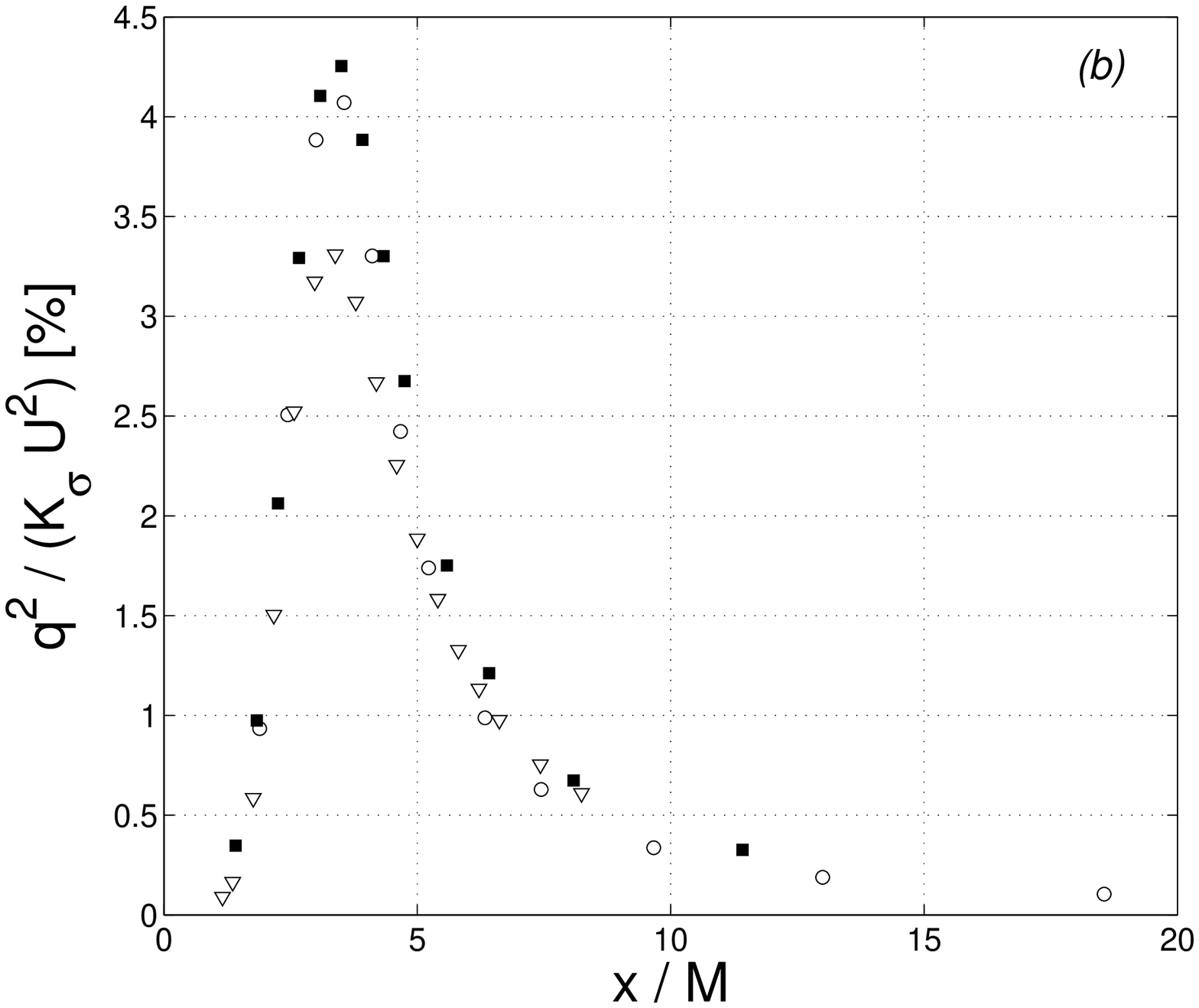}
\label{fig:GridsEnergyNorm}}
\caption{Streamwise evolution of \textit{(a)} the dimensionless integral length-scale
$\Lambda_u/M$ and \textit{(b)} the dimensionless turbulent kinetic energy
$q^2 / (K_\sigma U^2)$ measured on the wind-tunnel's centreline downstream various
perforated plates used separately. Symbols: $\bigcirc$, $D = 6mm$ and $M = 9mm$ ($Re = 1480$);
$\blacksquare$, $D = 8.5mm$ and $M = 12mm$ ($Re = 2100$); $\bigtriangledown$,
$D = 15mm$ and $M = 24.7mm$ ($Re = 3700$).}
\end{figure}

The mesh size $M_j$ controls the typical length-scale of energy-contained
structures generated downstream the plate $j$. This is emphasized by Figure
\ref{fig:IntegralScaleMonoNorm} showing the streamwise variation of the longitudinal integral
length-scale $\Lambda_u$ ($\equiv \int_0^\infty R_{uu}(h) dh$ with
$R_{uu}$ the longitudinal velocity autocorrelation coefficient and $h$
the spatial lag recovered from Taylor's hypothesis) obtained from \textit{LDV}
measurements, downstream from various individual plates. It is therefore reasonable to assimilate $M_j$ to the
preferential injection wavenumber $k$, i.e. $k \sim M_j^{-1}$.   
\vspace*{1ex}

In the present work, the smallest mesh size $M_1$ of the
\textit{MuSTI} corresponds to the Taylor micro-scale of the turbulence
generated by the \textit{MoSTI}, whilst the intermediate mesh size $M_2$
corresponds to a characteristic scale of its inertial range.

%\begin{table}[htbp]
%\begin{center}
%\begin{tabular}{ccccc}
%\hline
%$D (mm)$ &  $M (mm)$    &   $\sigma$    & $K_\sigma$ &  $Re$ \\
%\hline \hline
%$6$ &   $9$ &   $0.60$  &   $5.25$ & $1480$ \\
%$8.5$   &   $12$    &   $0.55$  &   $3.94$ & $2100$ \\
%$15$    &   $24.7$  &   $0.67$  &   $8.18$ & $3700$ \\
%\hline
%\end{tabular}
%\caption{Geometrical characteristics of individual perforated plates
%with $D$ the holes diameter, $M$ the mesh size  and $\sigma$ the
%blockage ratio. The parameter $K_{\sigma}$ is given by Eq.
%(\ref{eq:Ksigma}). Reynolds numbers $Re = U_\infty D / \nu$ are
%provided for $U_\infty = 3.7$ m/s.} \label{tab:Grids}
%\end{center}
%\end{table}

%\begin{table}
%\tbl{Geometrical characteristics of individual perforated plates}
%{\begin{tabular}{@{}ccccc}\toprule
%$D (mm)$ &  $M (mm)$    &   $\sigma$    & $K_\sigma$ &  $Re$$^{\rm a}$ \\
%\colrule
%$6$ &   $9$ &   $0.60$  &   $5.25$ & $1480$ \\
%$8.5$   &   $12$    &   $0.55$  &   $3.94$ & $2100$ \\
%$15$    &   $24.7$  &   $0.67$  &   $8.18$ & $3700$ \\
%\botrule
%\end{tabular}}
%\tabnote{$^{\rm a}$ $Re = U_\infty D / \nu$ with $U_\infty = 3.7$ m/s}
%\label{tab:Grids}
%\end{table}

\subsubsection{The plate spacing $L_j$}

The plate spacing $L_j$, which fixes the turbulence state, has to
be chosen carefully. Indeed, $L_j$ has to be neither too short,
in order to give time to turbulent structures generated by plate
$j$ to have strong enough turbulence passing through the plate $j+1$,
nor too long, to avoid the complete dissipation of those structures.
The evolution of the dimensionless turbulent kinetic energy $q^2$ 
($\equiv \left\langle u^2\right\rangle + \left\langle v^2\right\rangle
+\left\langle w^2\right\rangle$) measured with \textit{LDV}  
indicates that the best compromise is to set the plate $M_{j+1}$ at
the energy peak location of the plate $M_j$, i.e. $L_j \approx 3.5
M_j$ for this particular kind of plates (see Figure
\ref{fig:GridsEnergyNorm}). This distance has been therefore used
as the separation between two successive plates for the
\textit{MuSTI}.

\subsubsection{The pressure loss coefficient $K_{\sigma j}$}

Confronting the results of both Figures \ref{fig:IntegralScaleMonoNorm}
and \ref{fig:GridsEnergyNorm} suggests
that the individual energy spectrum $E_j(k)$ (defined by
$q^2 = \int_0^\infty E_j(k) dk$), induced by each
individual plate, can be written under the following form,
without loss of generality:

\begin{equation}
E_j(k) \sim K_{\sigma j} M_j g_j(k M_j)
\label{eq:spectrum}
\end{equation}

where $g_j(k M_j)$ is a dimensionless function representing the shape
of the energy spectrum in the wavenumber space. As
mentioned previously, the goal of the \textit{MuSTI} device
is to reproduce a cascade process which can be expressed through
a power-law:

\begin{equation}
E(k) = \sum_{j} E_j(k) \sim k^{-\gamma}
\label{eq:powerlaw}
\end{equation}

with $\gamma$ the power-law exponent. Combining Equations
(\ref{eq:spectrum}) and (\ref{eq:powerlaw}) leads therefore to:

\begin{equation}
K_{\sigma j} \sim M_j^{\gamma - 1}
\end{equation}

This relation implies therefore that the plates constituting the
multi-scale injector are self-similar. The diameter $D_j$ is then
simply obtained from the values of both $M_j$ and $K_{\sigma j}$
for each plate. In the present work, the power-law exponent $\gamma$
that we have used is equal to $1.57$ which is very close to the
well-known $5/3$ exponent reported for 3D turbulent flows
\citep{TennekesLumley1972}.

\subsubsection{The MuSTI device}

Following the steps mentioned hereinbefore, the complete design
of the \textit{MuSTI} device has been achieved. Its geometrical
properties are reported in Table \ref{tab:Injectors}.

\begin{table}[htbp]
\begin{center}
\begin{tabular}{cc|cc|cc|cc}
\hline
$D_1 (mm)$  &  3 & $M_1 (mm)$  &  4   & $\sigma_1$  & 0.49 &  $K_{\sigma 1}$  & 2.84\\
$D_2 (mm)$  &  6 & $M_2 (mm)$  &  9   & $\sigma_2$  & 0.60 &  $K_{\sigma 2}$  & 5.25\\
$D_3 (mm)$  & 15 & $M_3 (mm)$  & 24.7 & $\sigma_3$  & 0.67 &  $K_{\sigma 3}$  & 8.18\\
\hline
\end{tabular}
\caption{Geometrical characteristics of the multi-scale injector (\textit{MuSTI}).}
\label{tab:Injectors}
\end{center}
\end{table}

The direct comparison between the mono-scale and the multi-scale
injectors leads to a couple of remarks that it is important for the
reader to notice and keep in mind:

\begin{enumerate}
\item due to pressure loss addition, the total amount of
energy injected by the \textit{MuSTI} is roughly
twice that of the \textit{MoSTI} ($K_\sigma(\mbox{\textit{MuSTI}})
= \sum_j K_{\sigma j} = 16.27$).
Inverting the relation given in Equation (\ref{eq:Ksigma}) implies
that the equivalent blockage ratio of the \textit{MuSTI} injector
$\sigma(\mbox{\textit{MuSTI}})$ is about 13\% higher than that of
the \textit{MoSTI} device.
\vspace*{1ex}

\item for the \textit{MuSTI} injector, the genesis of the turbulence
starts about $45$mm ($= 3.5 M_1 + 3.5 M_2$) upstream from the wind
tunnel's inlet. As we show in the following, this property will strongly
impact the development of the turbulence downstream the injector.
\end{enumerate}

Moreover, we stress that care should be taken in extrapolating the results
we present in the following section as the design of the \textit{MuSTI}
has required to fix several degrees of freedom ($\gamma$ for instance). Indeed,
one can expect that those variables are of fundamental importance in
the development of turbulence downstream the multi-scale injector.
\vspace*{1ex}

In the following, for sake of simplicity, the largest mesh size
$M_3= 24.7$mm is used as a reference length-scale for both injectors.

\section{Results and discussion}

\subsection{The mean flow}

\begin{figure}[htbp]
\centering
\includegraphics[width=12cm]{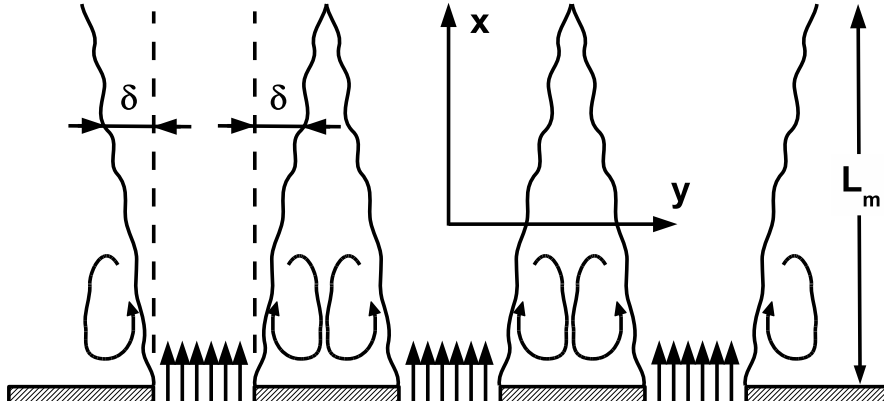}
\caption{Schematic representation of the near-field of an individual
perforated plate.} \label{fig:GridExit}
\end{figure}

In the near-field of individual perforated plates, jets issuing from
holes, extend via lateral spreading and merge at a distance $L_m$ from
the plate \citep{VillermauxHopfinger1994}. A schematic view of the
flow developing in the neighborhood of an individual perforated
plate is given in Figure \ref{fig:GridExit}.

\begin{figure}[htbp]
\centering
\subfigure
%{\includegraphics[width=7cm]{StreamMoSIG.eps}
{\includegraphics[width=7.5cm]{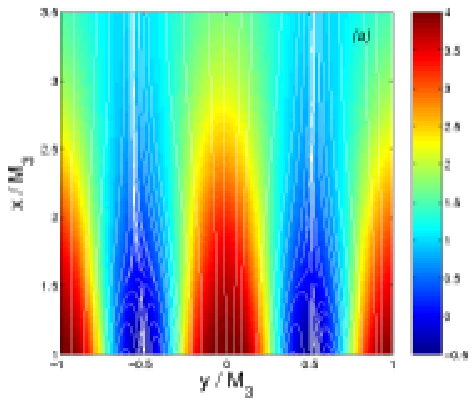}
\label{fig:StreamMoSIG}}  \subfigure
%{\includegraphics[width=7cm]{StreamMuSIG.eps}
{\includegraphics[width=7.5cm]{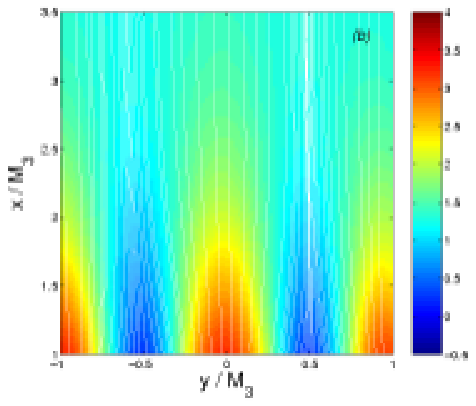}
\label{fig:StreamMuSIG}} \caption{Average streamlines and normalized
streamwise mean velocity map $U(x,y)/U_{\infty}$ (coloured
background) in the lee of \textit{(a)} the \textit{MoSTI} injector
and \textit{(b)} the \textit{MuSTI} injector computed from PIV
measurements.}
\end{figure}

This sketch compares very well with the map of scaled mean
velocity $U(x,y)/U_\infty$ in the lee of the \textit{MoSTI}
injector given in Figure \ref{fig:StreamMoSIG}. One can remark
the presence of recirculating bubbles before the jet merging
as evidenced by the average streamlines.

On the contrary, the average streamlines computed
for the \textit{MuSTI} injector (Figure \ref{fig:StreamMuSIG})
do not evidence recirculating bubbles between issuing jets, at least
downstream $x/M_3 = 1.15$ (we did not perform PIV measurements further
upstream because of laser light reflexions). This result is the
first illustration of the influence of the multi-scale
injection onto the flow development. Furthermore, one can
notice that the normalized mean velocity field $U(x,y) / U_\infty$
is quite different in intensity (i.e. colour) between the two
experiments (the same colorbar is used for both injectors).

\begin{figure}[htbp]
\centering 
\subfigure{\includegraphics[width=7.5cm]{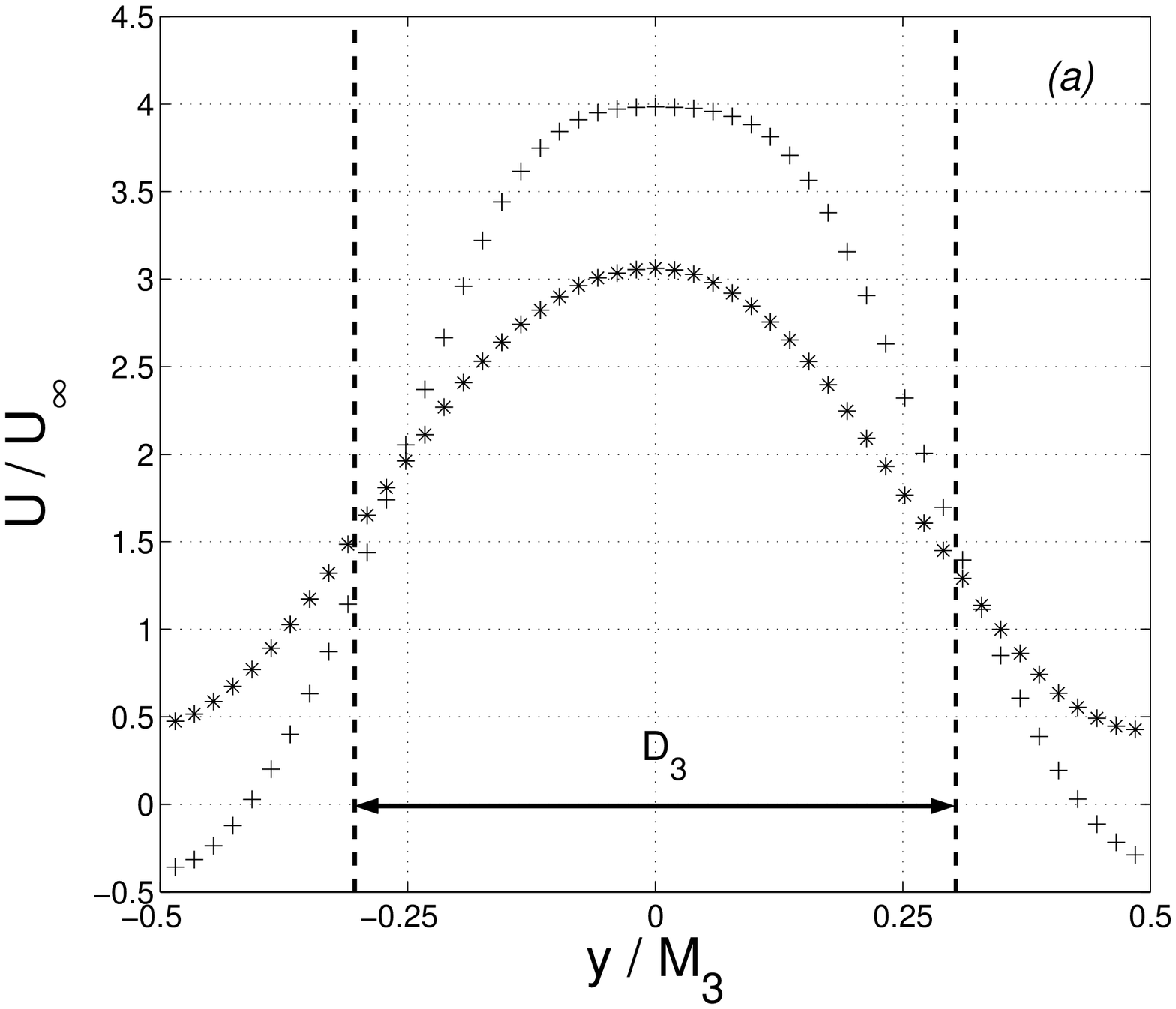}
\label{fig:MeanUExit}}
\subfigure{\includegraphics[width=7.5cm]{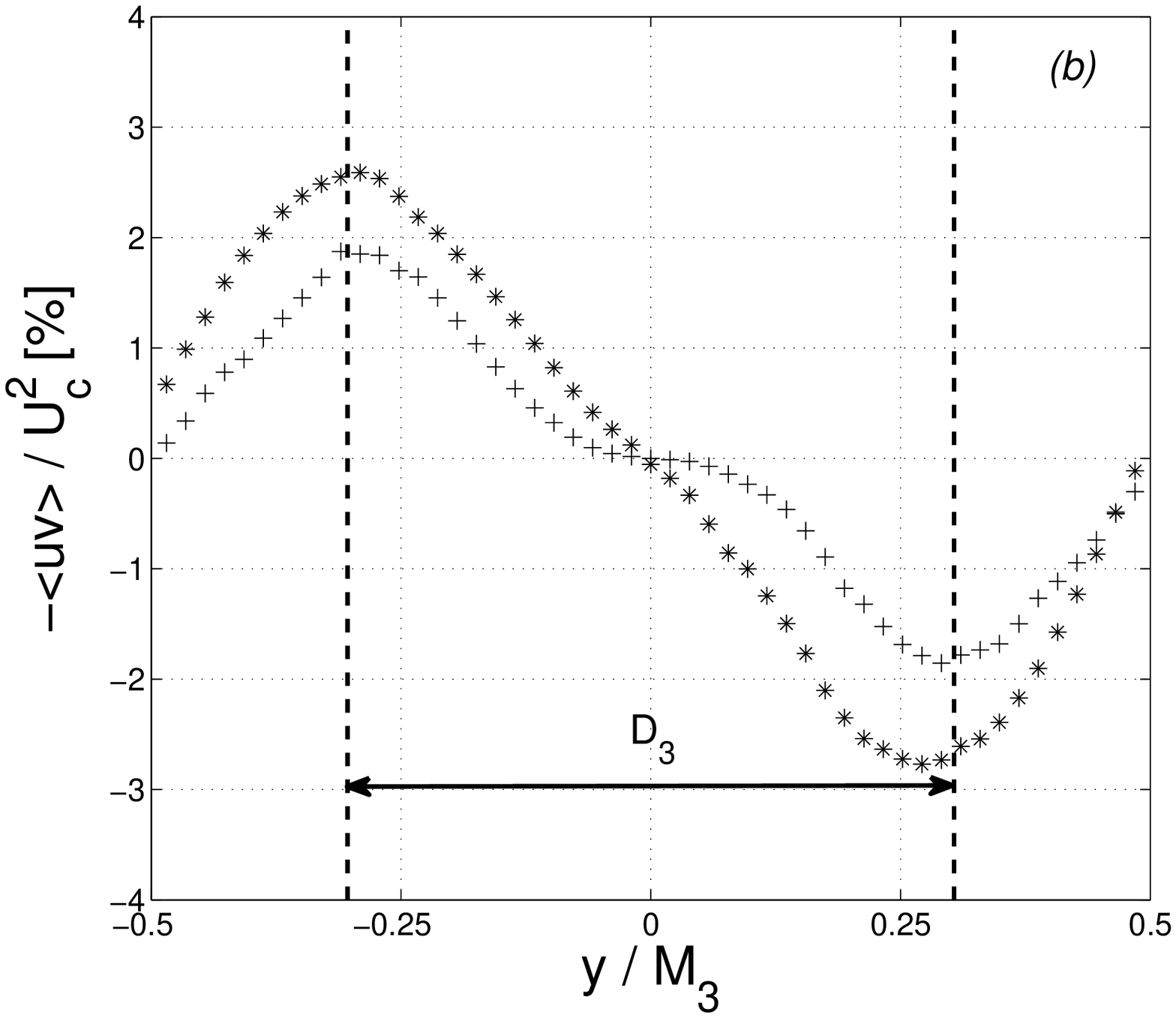}
\label{fig:uv}}
\caption{\textit{(a)} Dimensionless mean velocity profiles (central
hole) measured at $x/M_3 = 1.15$ with \textit{PIV}. \textit{(b)} 
Dimensionless transverse velocity flux measured
from \textit{PIV} at the same location for both
injectors. Symbols: \textbf{+}, \textit{MoSTI}
injector; \textbf{*}, \textit{MuSTI} injector.
The central hole position is illustrated by the vertical dashed lines.}
\end{figure}

This observation is highlighted by the spanwise profiles of $U/U_\infty$
plotted in Figure \ref{fig:MeanUExit} for $x/M_3 = 1.15$. The profile obtained
for the \textit{MoSTI} injector is characteristic of a potential core
(plateau around the centreline) surrounded by shear-layers. The latter
are responsible for the lateral jet spreading as shown by the jet
expansion compared to the hole's diameter. One can also remark that the
presence the recirculating bubbles induces negative velocities on
the jet boundaries.

The jet spreading downstream the \textit{MuSTI} injector
is clearly larger than that of the \textit{MoSTI}
injector. This behaviour reflects that the lateral shear-layer thickness
$\delta$ induced by the multi-scale injection is larger (by about $25$\%
at this location) than that of the reference device. Invoking mass-conservation
principle, this lateral expansion enhancement is followed by a decrease of the
centreline velocity $U_c = U(x, y=0, z=0)$ for the \textit{MuSTI} injector.
\vspace*{1ex}

This fast expansion of the shear-layers is related to the increase of the
transverse velocity flux $\left\langle uv\right\rangle$ as illustrated in
Figure \ref{fig:uv}. The jet spreading can be interpreted through the simple
model of turbulent viscosity $\nu_t$ which acts as an effective diffusion
coefficient:

\begin{equation}
\left\langle uv\right\rangle = -\nu_t \frac{\partial U}{\partial y}
\end{equation}

Due to the anticipated genesis of turbulence through small-scale
injection, the \textit{MuSTI} device accelerates the exchange of
energy in all directions increasing therefore the transverse fluxes
in comparison with the mono-scale injector. As a consequence, the
turbulent viscosity produced by the \textit{MuSTI} is about two times
bigger than that of the \textit{MoSTI}.

%%%%%%%%%%%%%%%%%%%%%%%%%%%%%%%%%%%%%%%%%%%%%%%%%%%%%%%%%%%%%%%%%%%%%%%%%%%%%%%%%%%%%%%%%%
%%%%%%%%%%%%%%%%%%%%%%%%%%%%%%%%%%%%%%%%%%%%%%%%%%%%%%%%%%%%%%%%%%%%%%%%%%%%%%%%%%%%%%%%%%
%%%%%%%%%%%%%%%%%%%%%%%%%%%%%%%%%%%%%%%%%%%%%%%%%%%%%%%%%%%%%%%%%%%%%%%%%%%%%%%%%%%%%%%%%%

\subsection{Homogeneity and isotropy}

One of the main objectives of the present work is to generate a turbulent
flow as nearly as possible homogeneous and isotropic. As shown hereinbefore,
the \textit{MuSTI} enhances the transverse velocity flux which are of
major importance for the redistribution of energy and by the way
for homogeneity and isotropy. It is obvious
from Figure \ref{fig:GridExit} that the homogeneity condition can only be
satisfied beyond the jet merging position symbolised by the length $L_m$.
This distance is straightly dependent on the shear-layers expansion rate
and is reached when $\delta / M_3 \sim 1$. The shear-layer
spreading results from the competition between convection and
turbulent diffusion characterised by the time-scales $\tau_{conv} \sim x / U$
and $\tau_{diff} \sim \delta^2/\nu_t$ respectively. Assuming that both
time-scales are of the same order of magnitude, it then comes:

\begin{equation}
\frac{\delta}{x} \sim Re_t^{-1/2}.
\end{equation}

with $Re_t = \frac{U x}{\nu_t}$ a turbulent Reynolds number. The jet
merging conditions ($\delta = M_3/2$ and $x = L_m$) leads therefore to:

\begin{equation}
\frac{M_3}{L_m} \sim \sqrt{\frac{\nu_t}{U L_m}}
\end{equation}

This equation illustrates the influence of the turbulent viscosity in
the jet merging phenomenon. The latter should therefore be accelerated
when the turbulent viscosity is increased.
\vspace*{1ex}

The jet merging distance $L_m$ has been experimentally estimated by comparing
the streamwise evolution of the relative
difference between the centreline mean velocity $U_c(x)$ and the minimum mean
velocity $U_{min}(x)$ (corresponding to the centre between 2 successive holes,
i.e. $y=M_3/2, z=0$). For this purpose, we introduce the dimensionless ratio
$H$ such that:

\begin{equation}
H = \frac{U_c - U_{min}}{U_c}
\end{equation}

\begin{figure}[htbp]
\centering \subfigure {\includegraphics[width=7.5cm]{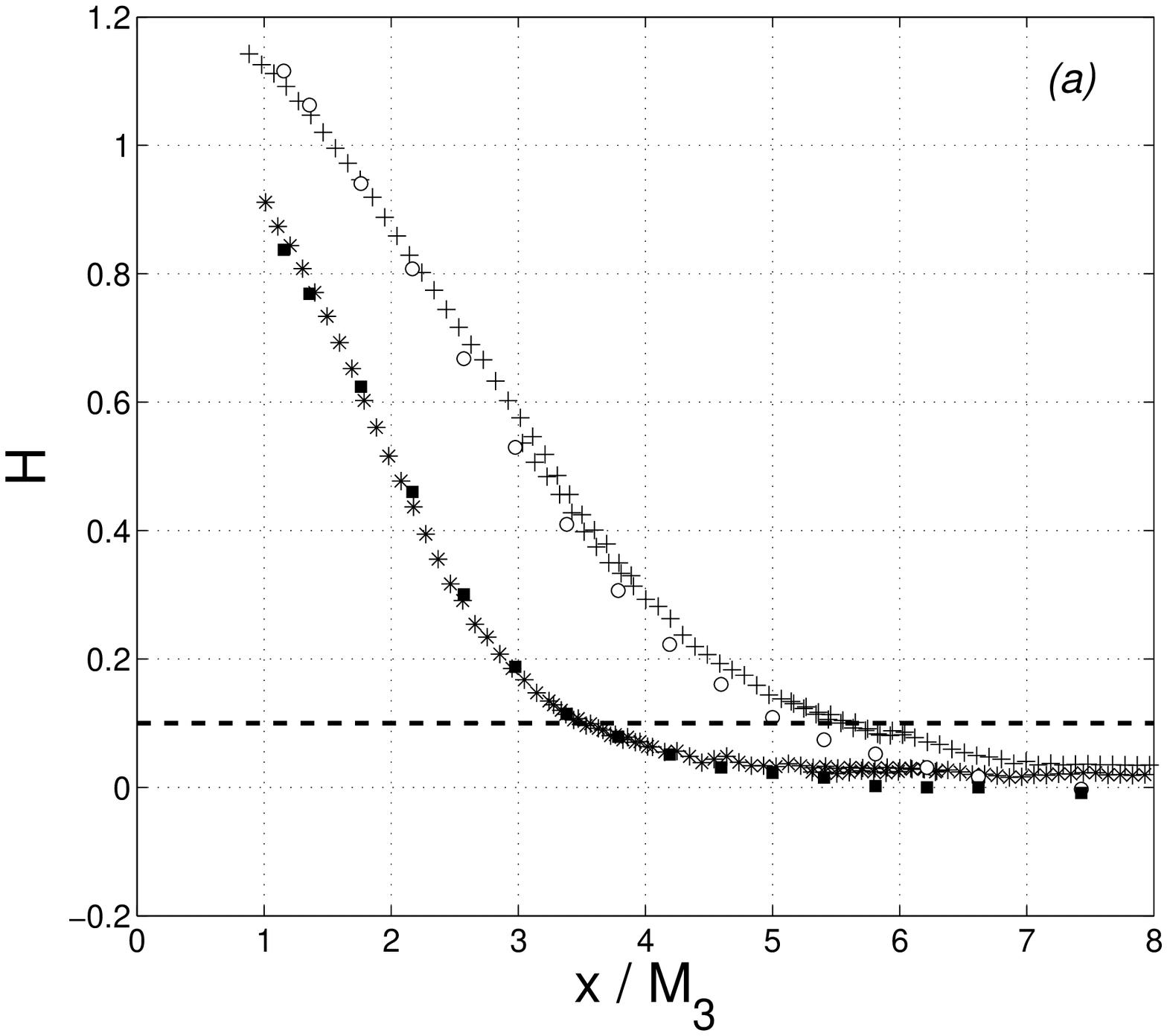}
\label{fig:Merging}}  \subfigure
{\includegraphics[width=7.5cm]{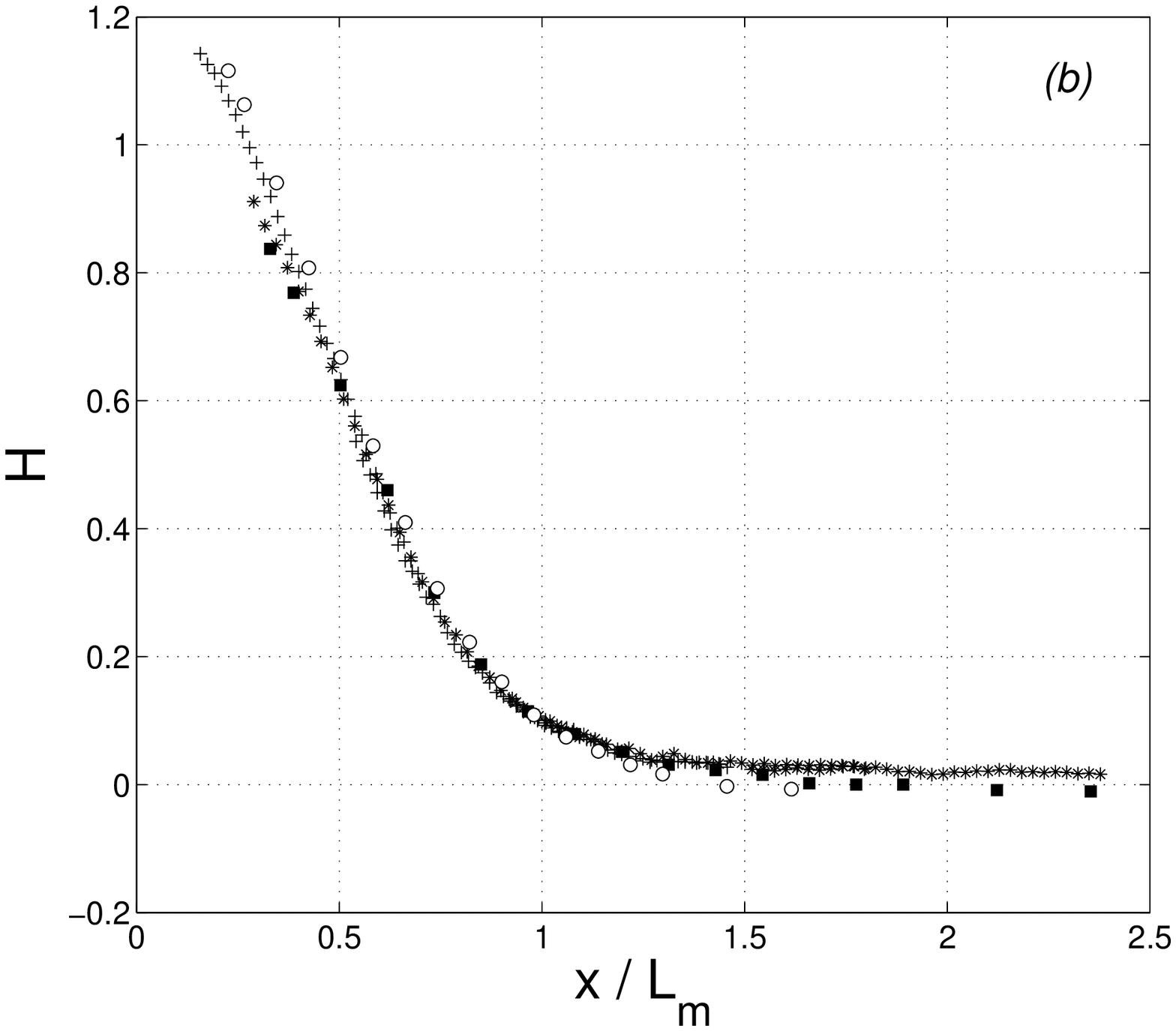}
\label{fig:MergingNorm}} \caption{Evolution of the ratio
$H$ as a function of the streamwise distance
$x$ scaled with the mesh size $M_3$ \textit{(a)} and with the merging
distance $L_m$ \textit{(b)}. The dashed line represents the merging
location $L_m$. Symbols: \textbf{+}, \textit{MoSTI} injector
(\textit{PIV} measurements); \textbf{*}, \textit{MuSTI}
injector (\textit{PIV} measurements); $\bigcirc$, \textit{MoSTI}
injector (\textit{LDV} measurements); $\blacksquare$, \textit{MuSTI}
injector (\textit{LDV} measurements).}
\end{figure}

It is worth to notice that $H$ is a representative criterion to evaluate the
flow homogeneity in the $y-z$ plane. Figure \ref{fig:Merging} shows the streamwise
variation of the ratio $H$ for both injectors. On this plot, the comparison of the
\textit{PIV} and \textit{LDV} measurements shows a very good agreement.

As expected, in the lee of the injectors, the high values of $H$
reflect the flow inhomogeneity in the transverse direction.
Further downstream, these inhomogeneities are
smoothed due to the shear-layer expansion resulting in the decrease of
$H$ which tends towards zero far away from
the plate. One can remark that the transverse inhomogeneity is
systematically smaller (tending therefore more rapidly towards zero)
for the \textit{MuSTI} injector. We define the merging distance $L_m$
as the position where the ratio $H$ becomes
smaller than $10$\%. This value is illustrated in Figure
\ref{fig:Merging} by the horizontal dashed line. The dimensionless
merging distance $L_m/M_3$ is found equal to $5.35 \pm 0.25$ and $3.50
\pm 0.02$ respectively for \textit{MoSTI} and \textit{MuSTI}
injectors. It is important to notice
that rescaling the streamwise distance $x$
by the merging length $L_m$ permits to collapse very well the ratio $H$
for both injectors as shown in Figure \ref{fig:MergingNorm}.
\vspace*{1ex}

\begin{figure}[htbp]
\centering \subfigure {\includegraphics[width=7.5cm]{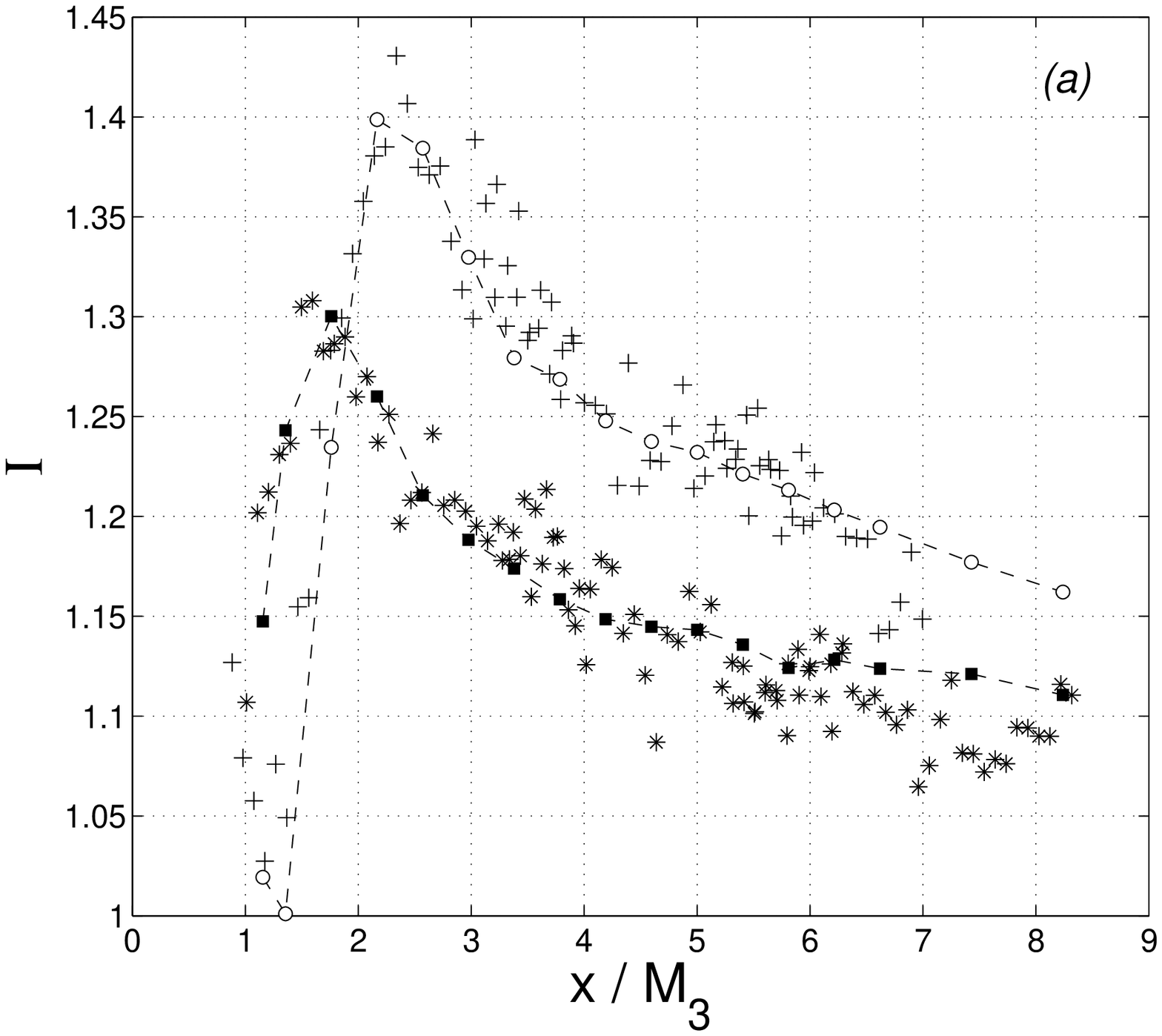}
\label{fig:Isotropy}}  \subfigure
{\includegraphics[width=7.5cm]{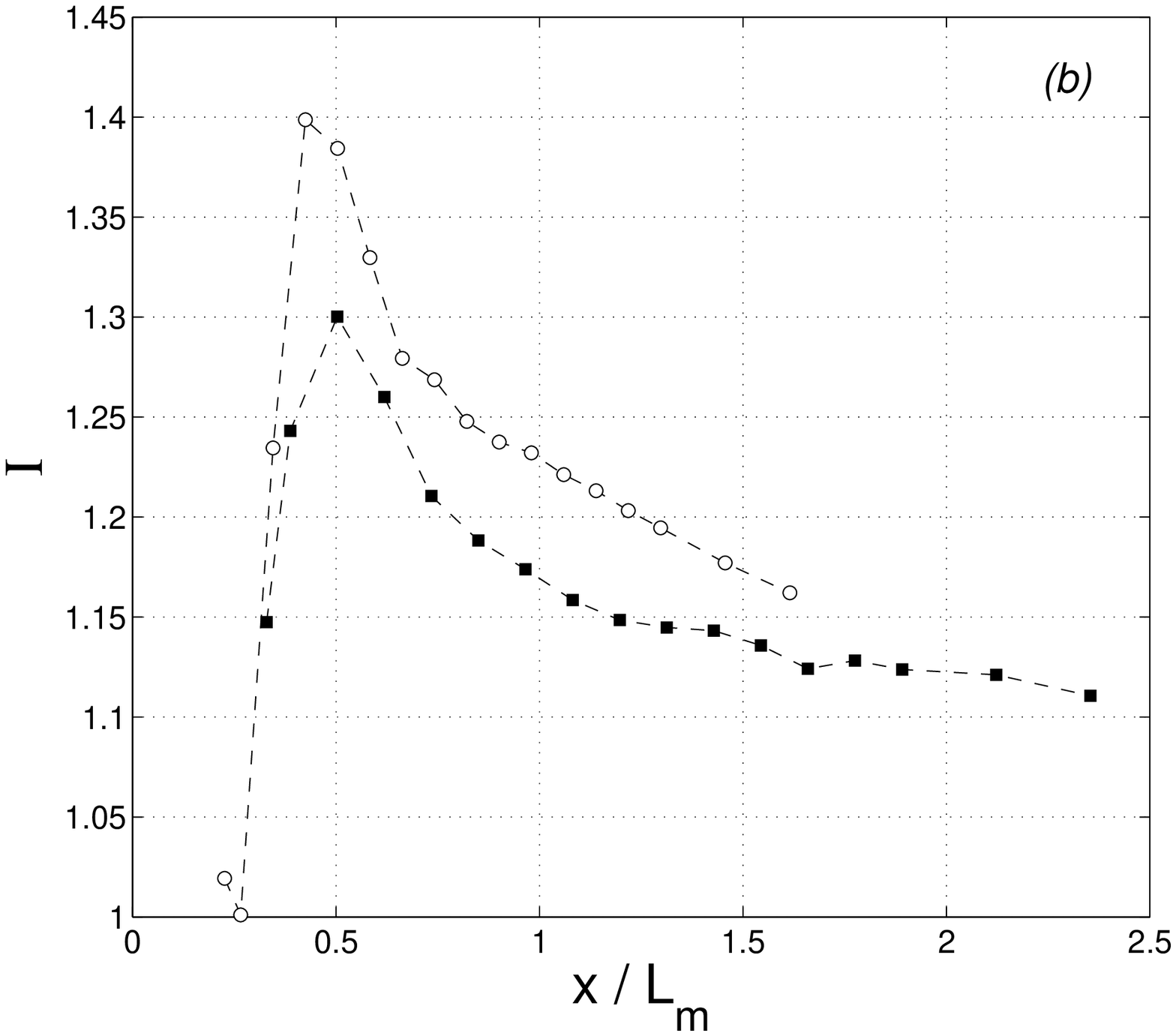}
\label{fig:IsotropyNorm}} \caption{Streamwise evolution of the
global isotropy factor $\sqrt{\left\langle u^2\right\rangle /
\left\langle v^2\right\rangle}$ with respect to the streamwise
distance $x$ scaled by the plate's mesh size $M_3$ \textit{(a)} and by
the merging length $L_m$ \textit{(b)}.
Symbols: \textbf{+},
\textit{MoSTI} injector (\textit{PIV} measurements);
\textbf{*}, \textit{MuSTI} injector (\textit{PIV}
measurements); $\bigcirc$, \textit{MoSTI} injector (\textit{LDV}
measurements); $\blacksquare$, \textit{MuSTI} injector (\textit{LDV}
measurements).}
\end{figure}

The global isotropy is commonly evaluated via the factor $I$ defined by:

\begin{equation}
I = \sqrt{\frac{\left\langle u^2\right\rangle}{\left\langle
v^2\right\rangle}}
\end{equation}

Figures \ref{fig:Isotropy} and \ref{fig:IsotropyNorm} show the streamwise
variation of the global isotropy ratio $I$ computed on the
tunnel's centreline for both injectors. As expected, the flow is 
strongly anisotropic close to the injector and then tends to
isotropy further away. The \textit{MuSTI} injector is characterised by
a better global isotropy ratio $I$ than the \textit{MoSTI} device.
Beyond $x/L_m = 1$, the global isotropy level obtained for the
\textit{MuSTI} injector is comparable to those reported by
Comte-Bellot \& Corrsin \citep{ComtebellotCorrsin1966} for
different kind of regular grids without secondary contraction.
Moreover, this level is better than the values reported by
Mydlarski \& Warhaft \citep{MW1996} for active grids ($\approx 1.2$)
and Hurst \& Vassilicos \citep{HurstVassilicos2007} for
fractal grids (between $\approx 1.2$ and $\approx 1.4$ depending
on the grid pattern).
\vspace*{1ex}

From these quantitative results three conclusions arise:

\begin{enumerate}
\item[(i)] the multi-scale injection strongly reduces the inhomogeneous
and anisotropic region downstream the injector. 
\item[(ii)] the nearly homogeneous and isotropic region appears much faster
(less than $5 M_3$) than in standard grid-generated turbulence ($40 M$ for
low-blockage regular grid for instance)
\item[(iii)] the merging length $L_m$ is revealed as a typical scaling length
for the two first order moments.
\end{enumerate}

\subsection{The one-point kinetic energy budget}

The second objective of our new multi-scale injector is to
generate intense turbulence level in the nearly homogeneous and
isotropic region. This property can be evaluated via the
dimensionless turbulent kinetic energy $q^2 / U_c^2$ measured on
the tunnel's centreline.

\begin{figure}[htbp]
\centering
\subfigure
{\includegraphics[width=7.5cm]{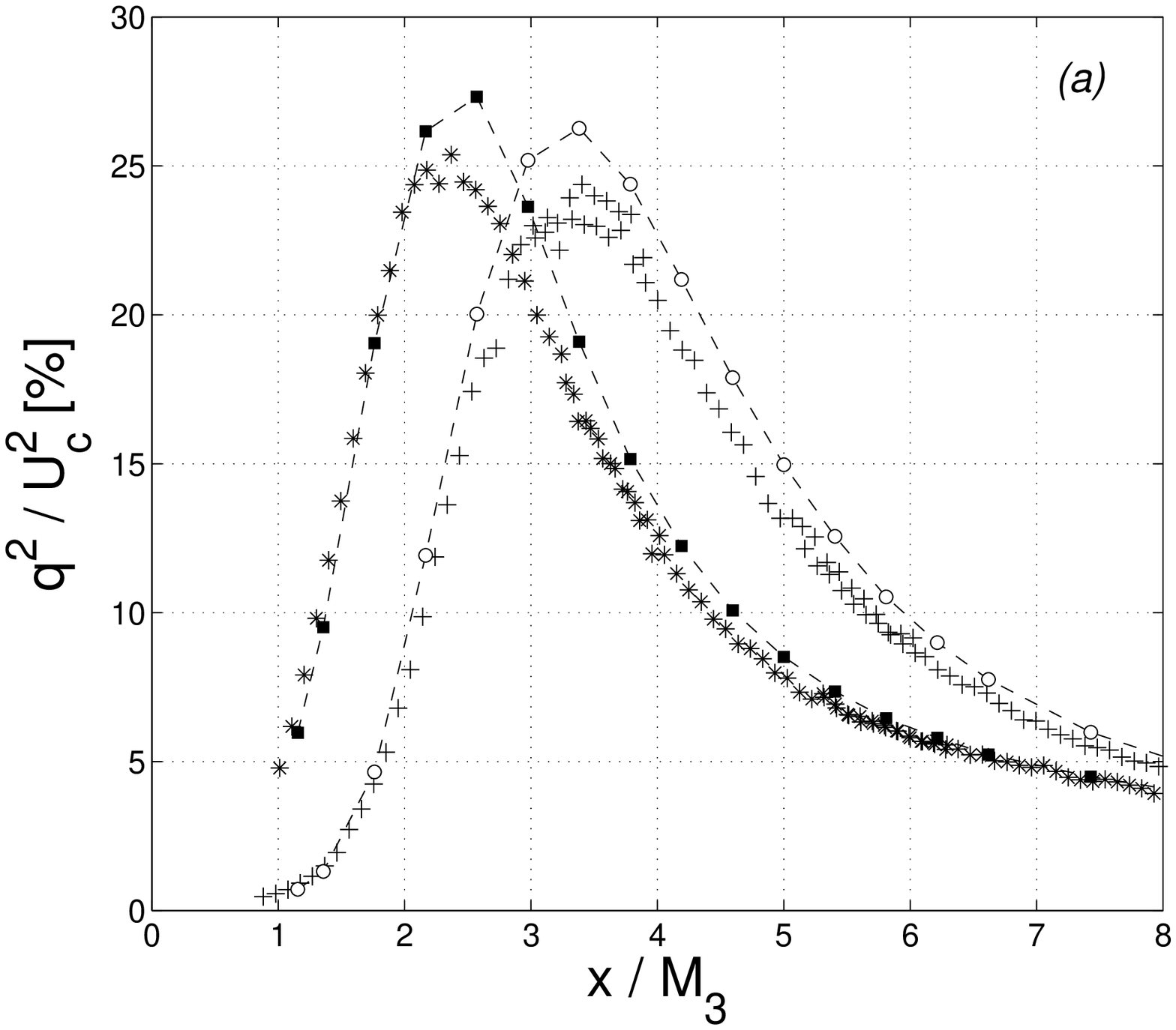}
\label{fig:Energy}}
\subfigure
{\includegraphics[width=7.5cm]{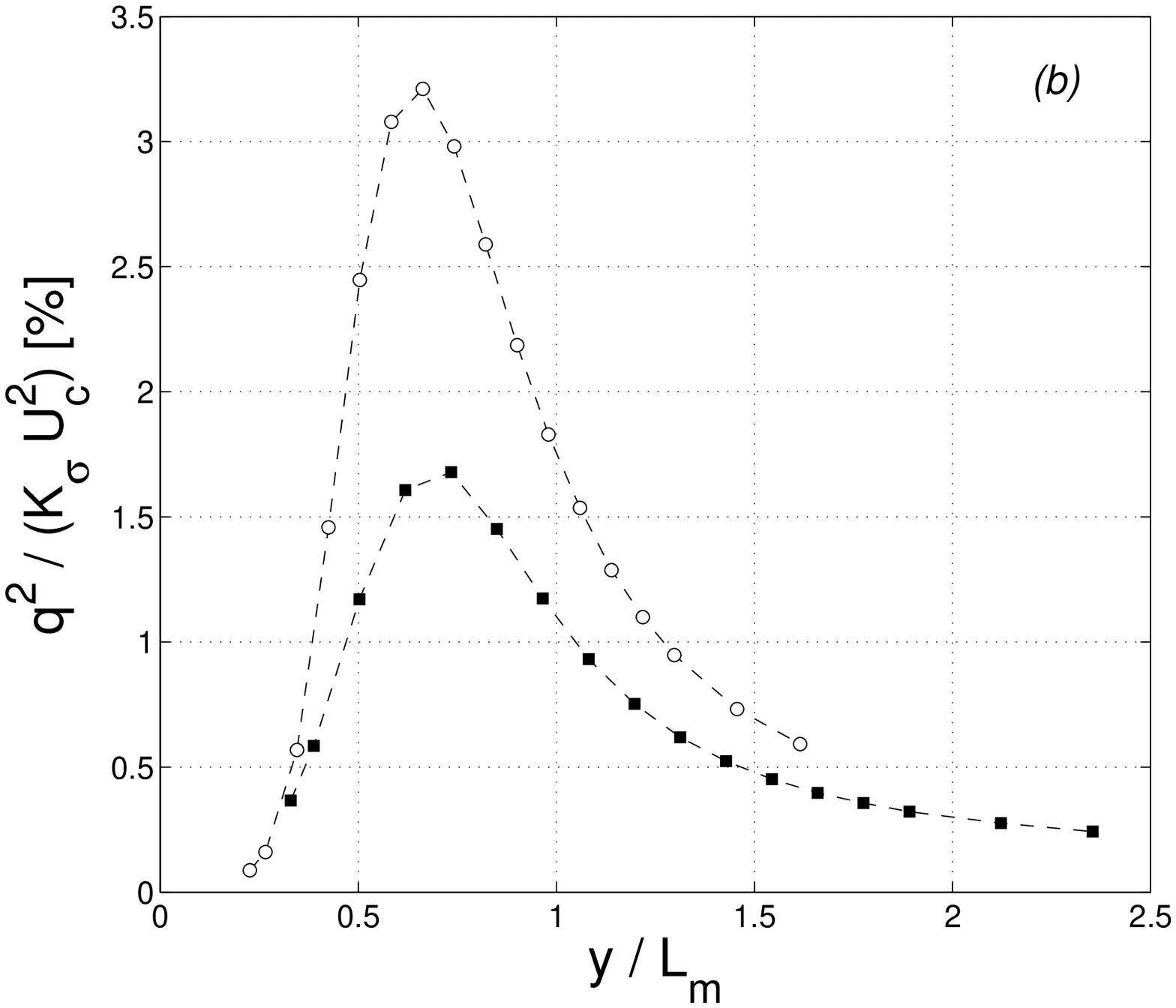}
\label{fig:EnergyNorm}}
\caption{\textit{(a)} Streamwise evolution of the dimensionless centreline turbulent kinetic energy $q^2 / U_c^2$ with respect to the streamwise distance $x$ scaled by the plate's mesh size $M_3$. \textit{(b)} Streamwise evolution of the dimensionless centreline turbulent kinetic energy $q^2 / (K_\sigma U_c^2)$ with respect to the streamwise distance $x$ scaled by the merging length $L_m$ . Symbols: \textbf{+}, \textit{MoSTI} injector (\textit{PIV} measurements); \textbf{*}, \textit{MuSTI} injector (\textit{PIV} measurements); $\bigcirc$, \textit{MoSTI} injector (\textit{LDV} measurements); $\blacksquare$, \textit{MuSTI} injector (\textit{LDV} measurements).}
\end{figure}

For both injectors, Figure \ref{fig:Energy}
shows the streamwise variation of this quantity measured from
both \textit{PIV} ($q^2 = \left\langle u^2\right\rangle +
2 \left\langle v^2\right\rangle$) and \textit{LDV}
($q^2 = \left\langle u^2\right\rangle + \left\langle v^2\right\rangle
+\left\langle w^2\right\rangle$). We point out that although the
experimental curves obtained from both techniques reasonably agree,
the \textit{PIV} measurements systematically
underestimate the turbulent kinetic energy compared to the
\textit{LDV} technique. This difference is explained by the low-pass
filtering of \textit{PIV} system \citep{Foucaultetal2004}. Small in
the near-field of the injectors because of the low turbulence level,
the discrepancies between \textit{PIV} and \textit{LDV} slightly
decrease far away from the plate due to the increase of the
Kolmogorov scale $\eta$ resulting into a better \textit{PIV}
resolution (see next section).

For both injectors, the turbulent kinetic energy builds up in the
vicinity of the latest plate until a maximum and then decays
monotonically. Due to transverse exchange enhancement, the evolution
of the turbulent kinetic energy $q^2$ is much faster for the
\textit{MuSTI} device than for the \textit{MoSTI} injector. This is
evidenced by the location of the peak of energy which appears much
closer to the injector for the \textit{MuSTI} device by about
$35$\%. Figure \ref{fig:EnergyNorm} evidences the strong dependence
between the energy peak location and the merging length $L_m$ meaning
that $q^2$ decays as soon as the jets issuing from the
plates interact. Moreover, this plot shows that, relatively
to the total pressure loss imposed by the injectors, the turbulent
kinetic energy generated by the \textit{MuSTI} decays much more
slowly than that of the \textit{MoSTI}. This is a direct consequence of
the acceleration of the transverse exchanges due to the multi-scale
injection.
\vspace*{1ex}  

Derived from Navier-Stokes equations \citep{Pope2000}, the governing equation
of turbulent kinetic energy $q^2$ can be expressed, in cylindrical-coordinates,
as follows:

\begin{eqnarray}
-\underbrace{U \frac{\partial q^2}{\partial x} - V \frac{\partial
q^2}{\partial y}}_{\mbox{Convection}} - \underbrace{\frac{\partial
\left\langle u q^2\right\rangle}{\partial x} - \frac{1}{y}
\frac{\partial y \left\langle v q^2\right\rangle}{\partial
y}}_{\mbox{Diffusion}} \nonumber \\
- \underbrace{2 \left[\left\langle u^2\right\rangle
\frac{\partial U}{\partial x} + \left\langle v^2\right\rangle
\frac{\partial V}{\partial y} + \left\langle uv\right\rangle
\frac{\partial U}{\partial y}\right]}_{\mbox{Production}} \nonumber \\
+P - 2 \epsilon = 0 \label{eq:budget}
\end{eqnarray}

where $P$ represents the turbulent kinetic energy transport
by pressure and $\epsilon$ is the
turbulent kinetic energy dissipation rate per mass unit. The
first three  terms (Convection, Diffusion and Production)
of Equation (\ref{eq:budget}) can be easily evaluated
from \textit{PIV} measurements. The pressure term $P$ is usually
unaccessible by a direct evaluation and is commonly deduced from the
budget of Equation (\ref{eq:budget}). Unfortunately, low-pass
filtering inherent to \textit{PIV} system avoids the accurate
estimation of the dissipation $\epsilon$
(see e.g. \citep{Foucaultetal2004}, \citep{Lavoieetal2007b}). The
spectral corrections recently given by Lavoie \& al.
\citep{Lavoieetal2007b} are restricted to homogeneous and isotropic
turbulence which is obviously not representative of the injector
near-field. Using the continuity equation and dropping off small
terms, Equation (\ref{eq:budget}) simplifies to the following form on the
tunnel's centreline:

\begin{equation}
-U_c \frac{\partial q^2}{\partial x} - \frac{1}{y} \frac{\partial y
\left\langle v q^2\right\rangle}{\partial y} - 2 \left(\left\langle
u^2\right\rangle -  \left\langle v^2\right\rangle\right)
\frac{\partial U_c}{\partial x} + P - 2 \epsilon
= 0. \label{eq:budget2}
\end{equation}

\begin{figure}[htbp]
\centering
\subfigure
{\includegraphics[width=7.5cm]{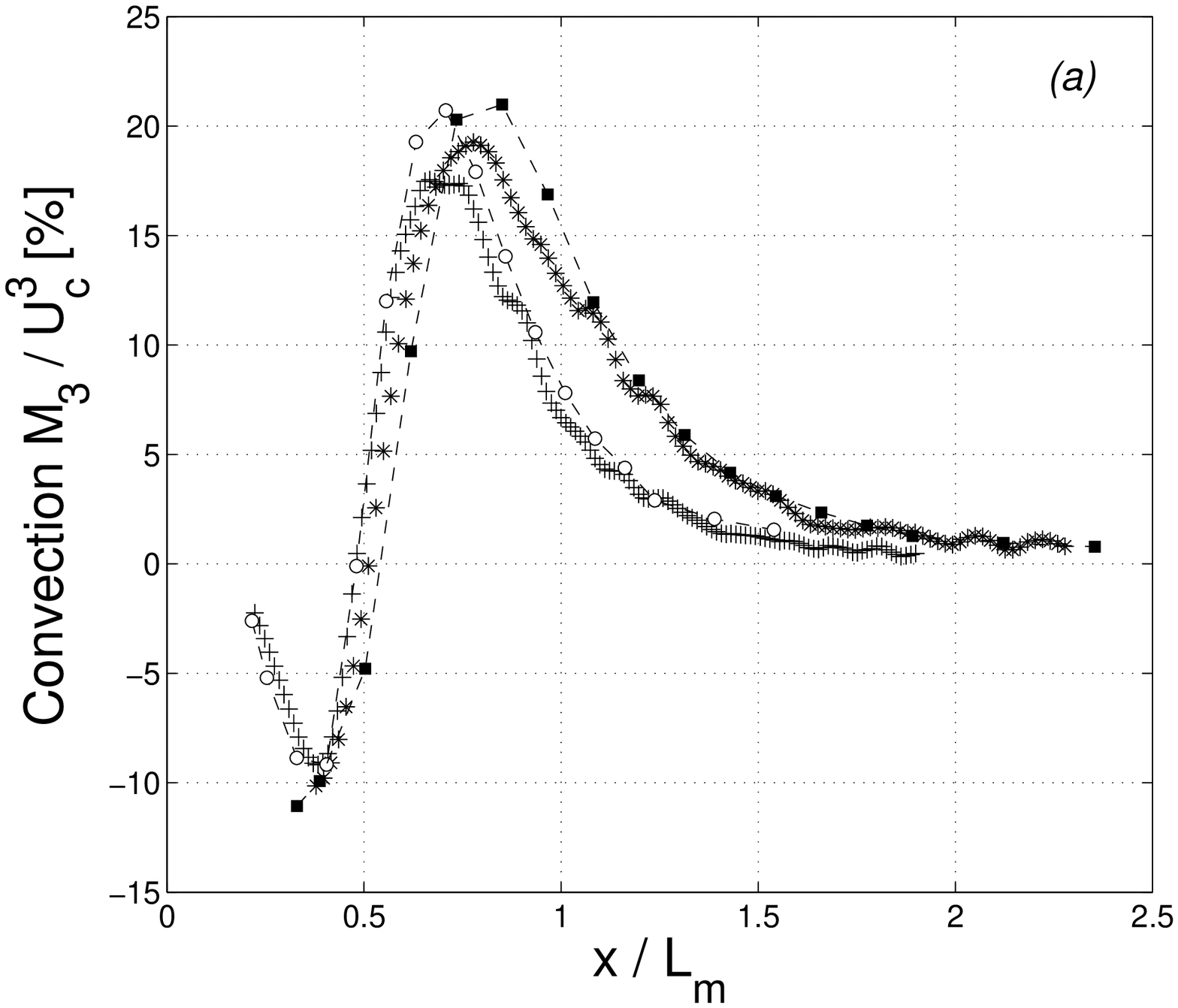}
\label{fig:Convection}}
\subfigure
{\includegraphics[width=7.5cm]{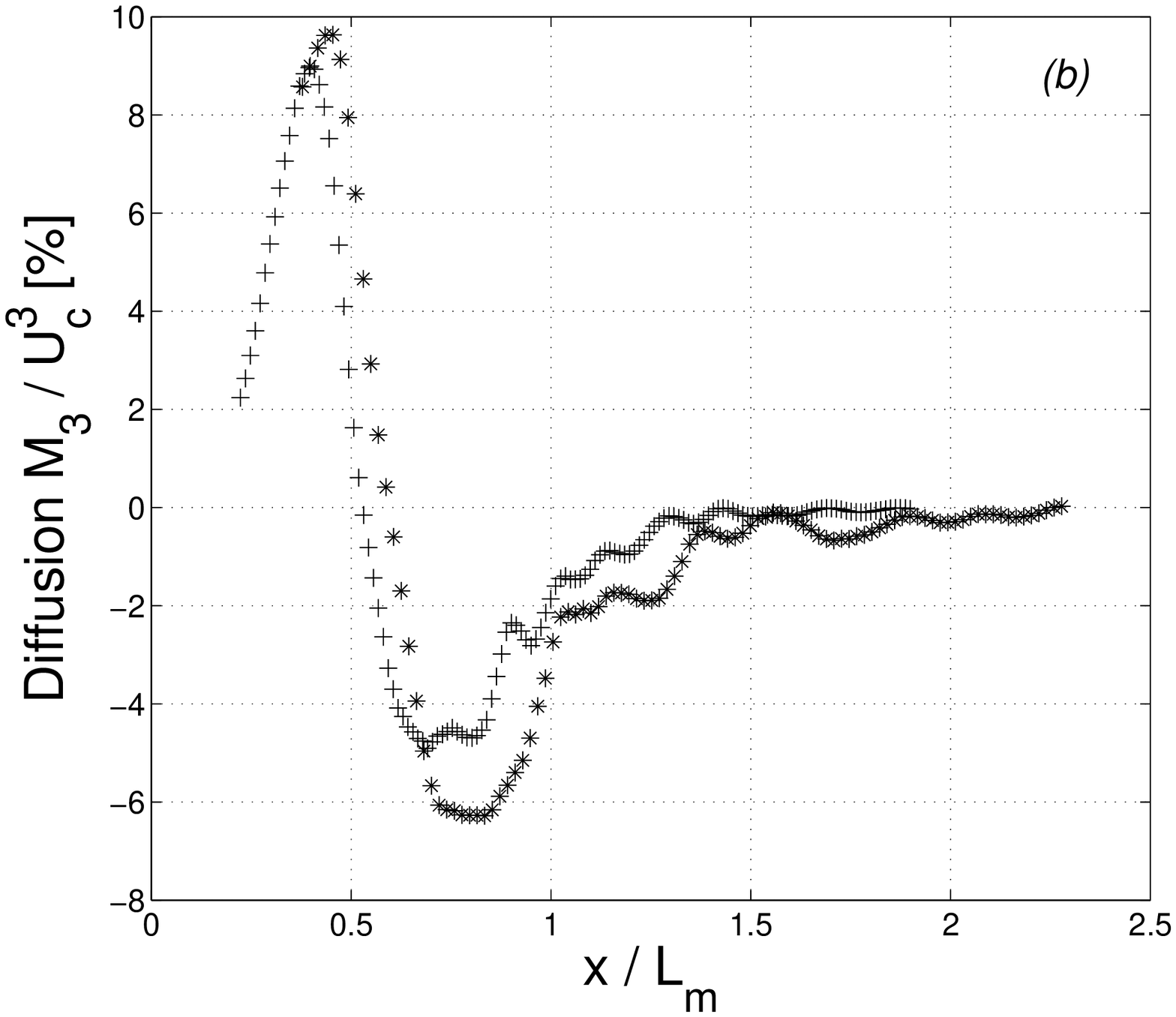}
\label{fig:Diffusion}}
\vspace{0.1cm}
\subfigure
{\includegraphics[width=7.5cm]{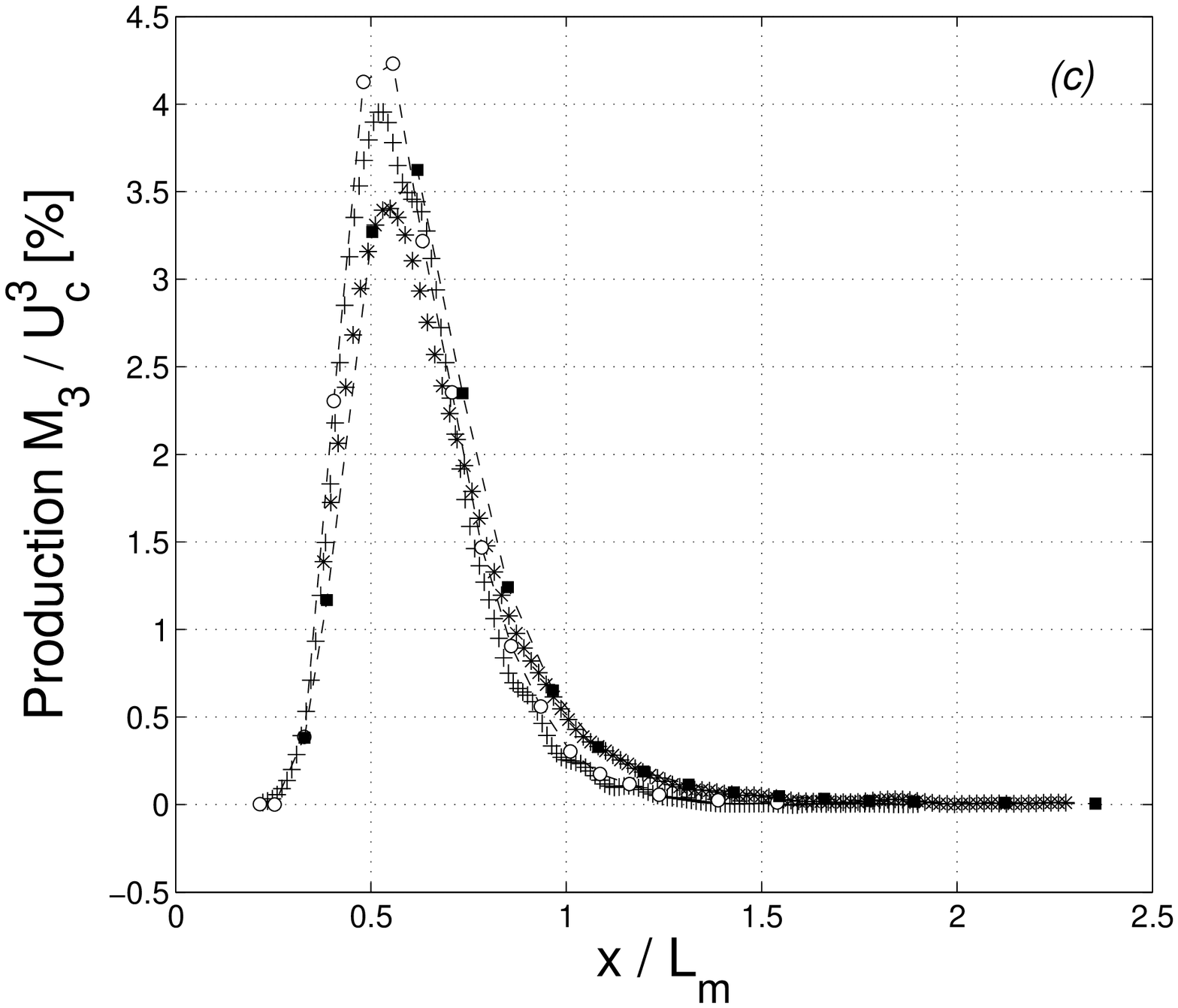}
\label{fig:Production}}
\subfigure
{\includegraphics[width=7.5cm]{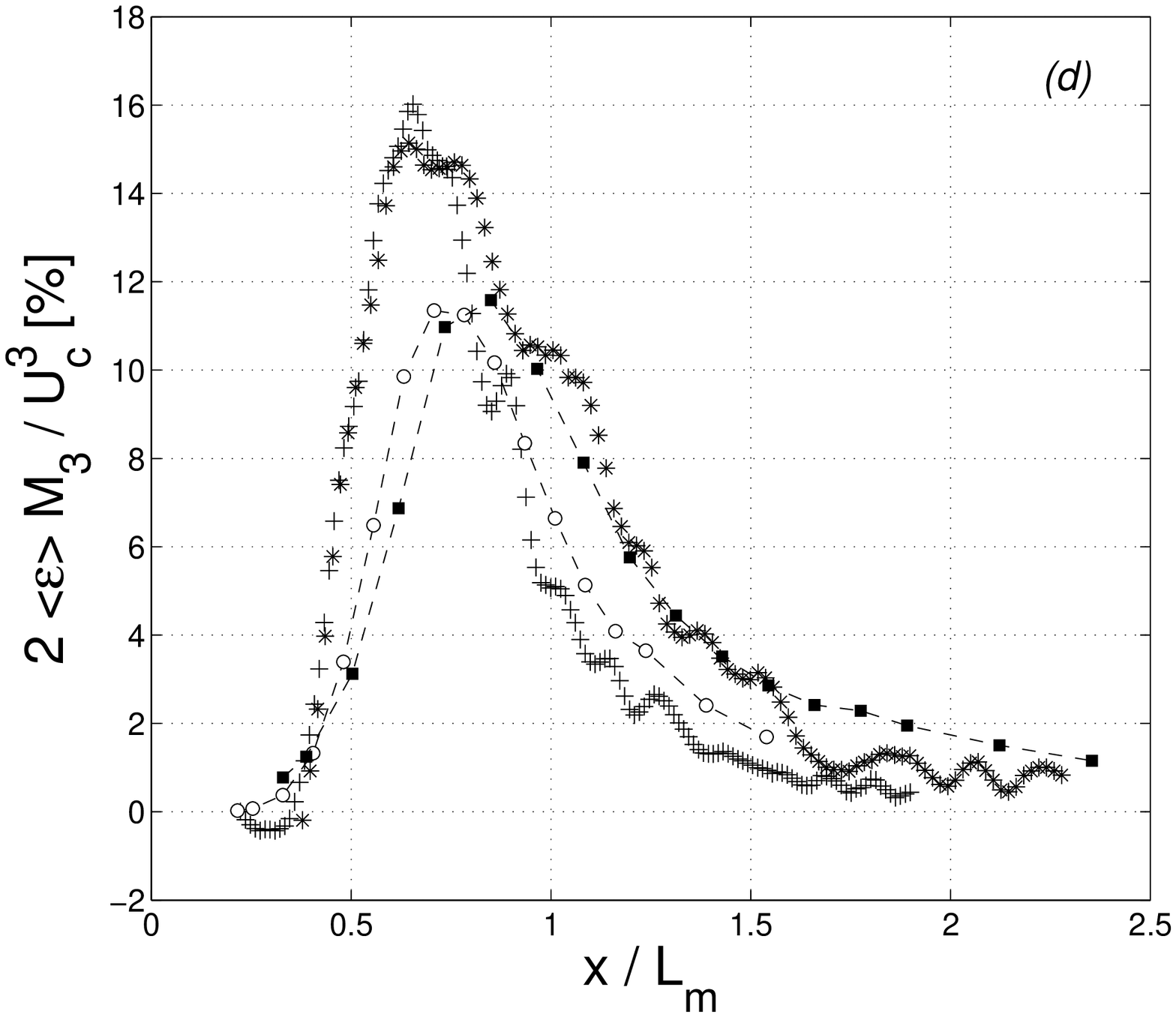}
\label{fig:Dissipation}}
\caption{Streamwise evolution of the terms appearing in the budget equation of the turbulent kinetic energy: Convection \textit{(a)}, Diffusion \textit{(b)}, Production \textit{(c)} and Dissipation \textit{(d)} with respect to the normalized streamwise distance $x/L_m$ for both injectors. Symbols: \textbf{+}, \textit{MoSTI} injector (\textit{PIV} measurements); \textbf{*}, \textit{MuSTI} injector (\textit{PIV} measurements); $\bigcirc$, \textit{MoSTI} injector (\textit{LDV} measurements); $\blacksquare$, \textit{MuSTI} injector (\textit{LDV} measurements).}
\end{figure}

The different terms involved in Equation (\ref{eq:budget2}) are plotted
in Figures \ref{fig:Convection}, \ref{fig:Diffusion},
\ref{fig:Production} and \ref{fig:Dissipation} with respect to the
dimensionless distance $x/L_m$ for both
injectors. The dissipation $\epsilon$ is
estimated by two independent ways:

\begin{equation}
\epsilon_{q} = - \mbox{Convection} - \mbox{Diffusion} -
\mbox{Production}, \label{eq:DissipPIV}
\end{equation}

and

\begin{equation}
\epsilon_{h} = 3 \nu \left[\left\langle
\left(\frac{\partial u}{\partial x}\right)^2\right\rangle +
\left\langle \left(\frac{\partial v}{\partial
x}\right)^2\right\rangle + \left\langle \left(\frac{\partial
w}{\partial x}\right)^2\right\rangle\right],  \label{eq:DissipLDV}
\end{equation}
where gradients are inferred from \textit{LDV} measurements and Taylor's
hypothesis.

Equation (\ref{eq:DissipPIV}) evaluates dissipation through
large-scale properties (\textit{PIV}), while Equation (\ref{eq:DissipLDV}) is a more
direct evaluation through small-scale gradients (\textit{LDV}) for which
homogeneity is assumed \citep{Danailaetal2002}. The estimation of
$\epsilon_{q}$ requires the pressure-related
term $P$ to be negligible. This assumption can be checked by
comparing both $\epsilon_{q}$ and
$\epsilon_{h}$. Results plotted in Figure
\ref{fig:Dissipation} show that this assumption is fairly well
fulfilled beyond $x/L_m > 1$, i.e. in the nearly homogeneous and
isotropic region.

Although the various terms of the budget equation evolve in a very
similar way for both injectors, some differences remain, especially
beyond $x/L_m = 0.5$. Upstream this position, the lateral diffusion
term almost compensates the axial convection term which contributes
negatively to the turbulent kinetic energy budget meaning that the
mean flow convects energy from low energy regions towards high
energy regions. On the contrary, the lateral energy diffusion brings
energy produced in the lateral shear layer towards the centreline.

Beyond $x/L_m = 0.5$, convection and diffusion terms change sign. This
position coincide with the end of the potential core ($\delta / D_3
\sim 1$) resulting into a peak of turbulent kinetic energy
production which is compensated by the negative lateral diffusion
term. Then, these two terms decrease in magnitude to become
negligible further downstream implying
\citep{ComtebellotCorrsin1966}:

\begin{equation}
-U_c \frac{\partial q^2}{\partial x} - 2 
\epsilon = 0. \label{eq:budget3}
\end{equation}

This equation reflects that in decaying turbulence, the convection
term is compensated by the dissipation. In this region, all
terms computed for the \textit{MuSTI} injectors are higher in magnitude
than those of the \textit{MoSTI} device.

The results obtained from the study of the turbulent kinetic energy
lead to conclusions:

\begin{enumerate}
\item[(i)] the turbulent energy generated by the multi-scale injection
is more intense partly due to an higher pressure loss.

\item[(ii)] the nearly homogeneous and isotropic region is characterised
by a very large turbulence intensity ($\sqrt{q^2 / 3}/U_c \approx 15\mbox{\%}$)
compared to standard grid-generated turbulence ($\approx 3\mbox{\%}$).
\end{enumerate}

\subsection{Characteristic length-scales and two-point kinetic energy budget}

\begin{figure}[htbp]
\centering
\subfigure
{\includegraphics[width=7cm]{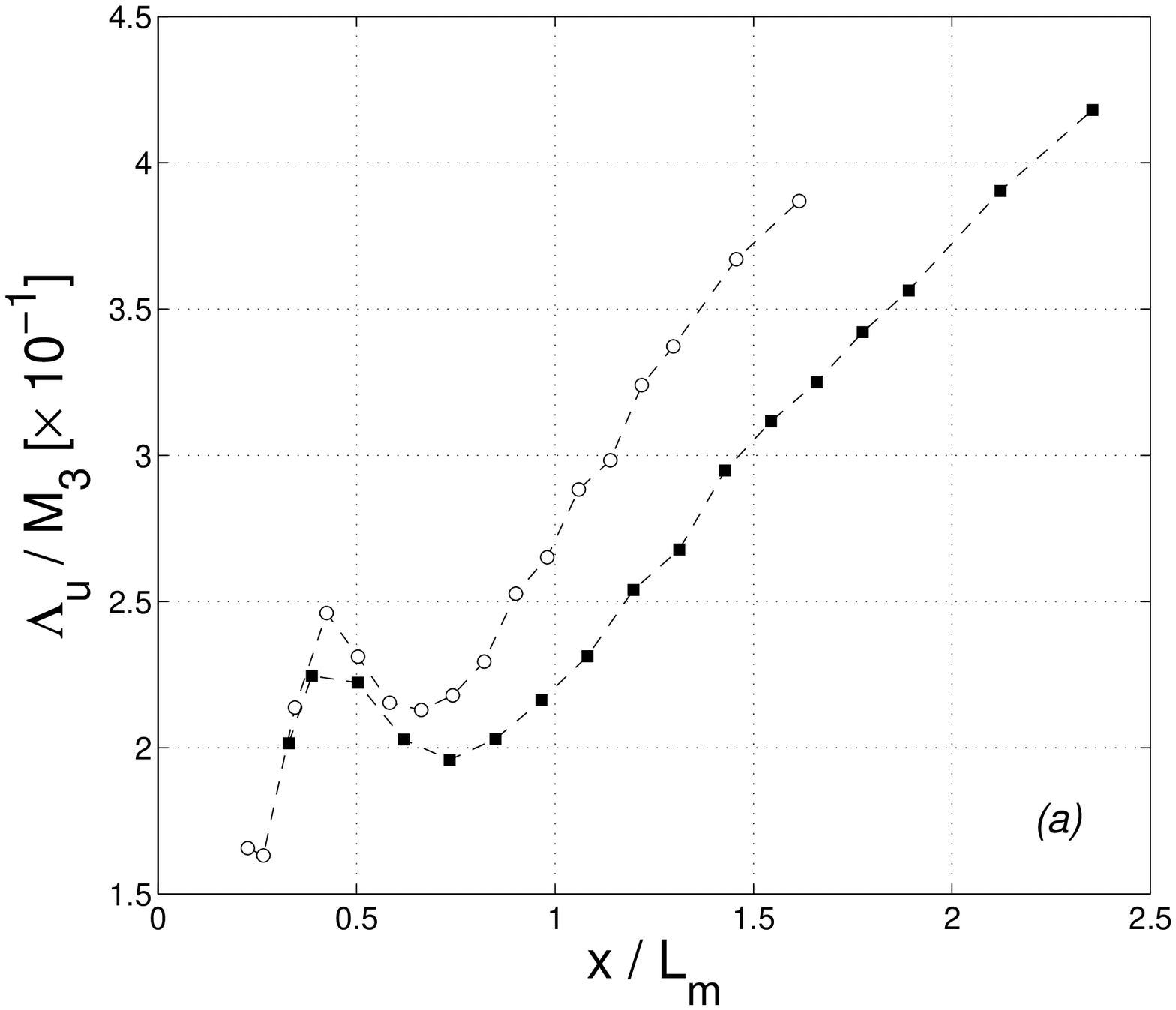}
\label{fig:IntScaleNorm}}
\subfigure
{\includegraphics[width=7cm]{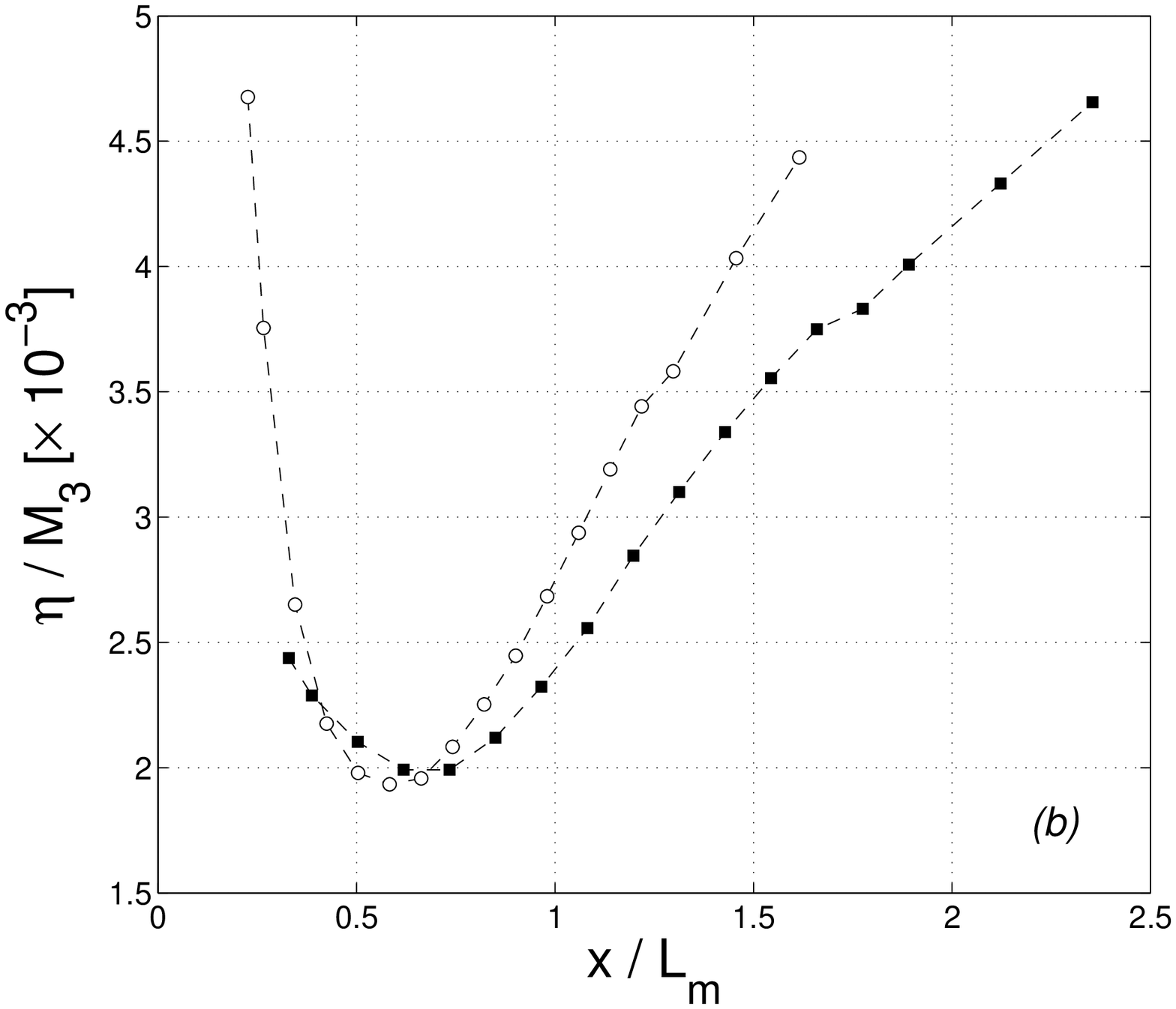}
\label{fig:KolmogorovNorm}}
\caption{Streamwise variation of the normalized integral length-scale
$\Lambda_u / M_3$ \textit{(a)} and Kolmogorov scale $\eta / M_3$ \textit{(b)}.
Symbols: $\bigcirc$, \textit{MoSTI} injector; $\blacksquare$, \textit{MuSTI} injector.}
\end{figure}

We remind that the main goal of the present work is to develop
a new kind of turbulence injector enable to quickly inject energy
over a broad range of scales. The implicit idea is to act onto
energy transfer by modifying the energy's cascade. The latter
is characterised by both the integral length-scale $\Lambda_u$
(computed from the auto-covariance of streamwise fluctuation $u$)
and the Kolmogorov scale $\eta$ ($\equiv (\nu^3 / \epsilon_h)^{1/4}$)
which are plotted in dimensionless form in
Figures \ref{fig:IntScaleNorm} and \ref{fig:KolmogorovNorm} for
both injectors on the tunnel's centreline. In the nearly homogeneous
and isotropic region, $\Lambda_u$  and $\eta$ increase with respect to
$x$. For both injectors, the integral length-scale $\Lambda_u$ is
controlled by the largest mesh size $M_3$ which can be interpreted as
the spanwise interaction distance. However, the integral length-scale
$\Lambda_u$ generated by the \textit{MuSTI} injector is significantly
smaller than that of the \textit{MoSTI} injector. This observation
applies also to the Kolmogorov scale.

\begin{figure}[htbp]
\centering
\includegraphics[width=10cm]{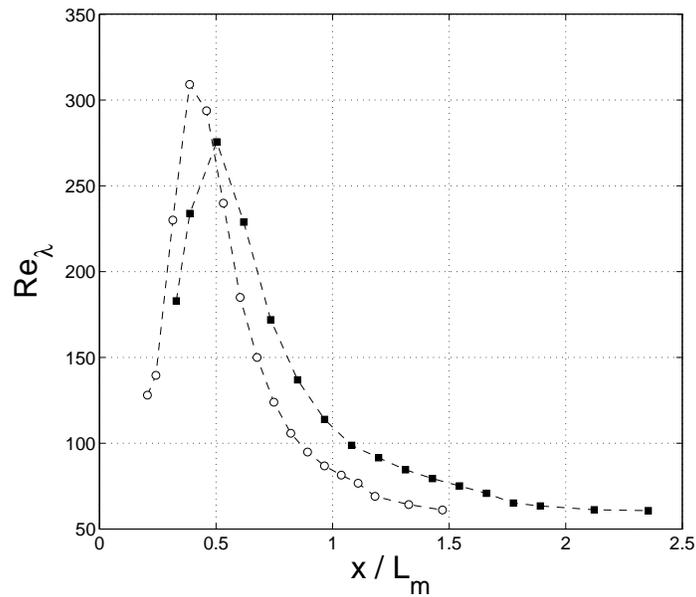}
\caption{Taylor-based Reynolds number evolution. Symbols: $\bigcirc$, \textit{MoSTI} injector;
$\blacksquare$, \textit{MuSTI} injector.}
\label{fig:Rlambda}
\end{figure}

Although the turbulent scales generated by the \textit{MuSTI} device
are systematically smaller than those of the \textit{MoSTI}
injector, the Taylor-based Reynolds number $Re_\lambda$ ($\equiv
\sqrt{q^2 / 3} \lambda / \nu$ with $\lambda = \sqrt{5 \nu q^2 /
\epsilon_h}$
the Taylor micro-scale) of the \textit{MuSTI} injector is
higher (by about $30$\%) than that of the \textit{MoSTI} injector
as shown in Figure \ref{fig:Rlambda}. This is
due to the acceleration of cascade exchange allowing to reach
isotropic region with much larger turbulence intensity which
compensates the decrease in turbulent length-scales. Although the
values of $Re_\lambda$ reported here remain moderate ($\approx 80$),
one has to keep in mind that the experimental wind tunnel is quite
small in comparison with usual facilities (see e.g.
\citep{ComtebellotCorrsin1966} for comparison).

The way the turbulent energy is injected by both devices can
be investigated with structure functions which represent the energy
of a scale.
These are defined by using spatial increments, i.e. velocity differences
between two space points separated by a vector $\bm{r}$:

\begin{equation}
\Delta u_i (\bm{r}) = u_i(\bm{x} + \bm{r}) - u_i(\bm{r})
\end{equation}

where $i$ designates any velocity component. Under the isotropy
hypothesis, the scalar $S_{2q}=(\Delta q)^2 (\bm{r}) \equiv (\Delta
u_i)^2 (\bm{r})$ only depends on $r$, the modulus of the separation
$\bm{r}$, and represents the total kinetic energy of the scale $r$.
These increments are calculated using LDV data, and the separation
$r$ is obtained from the temporal lag, via the Taylor's hypothesis
(the local Taylor's hypothesis has also been used, with no significant
difference on the results). An important remark is to be done here.
The structure functions calculation from data non-uniformly sampled
in time (\textit{LDV} technique) does not require an equidistant time-resampling (as it is
the case for calculating spectra) unlike Fourier transform.

\begin{figure}[htbp]
\centering
\subfigure {\includegraphics[width=7.5cm]{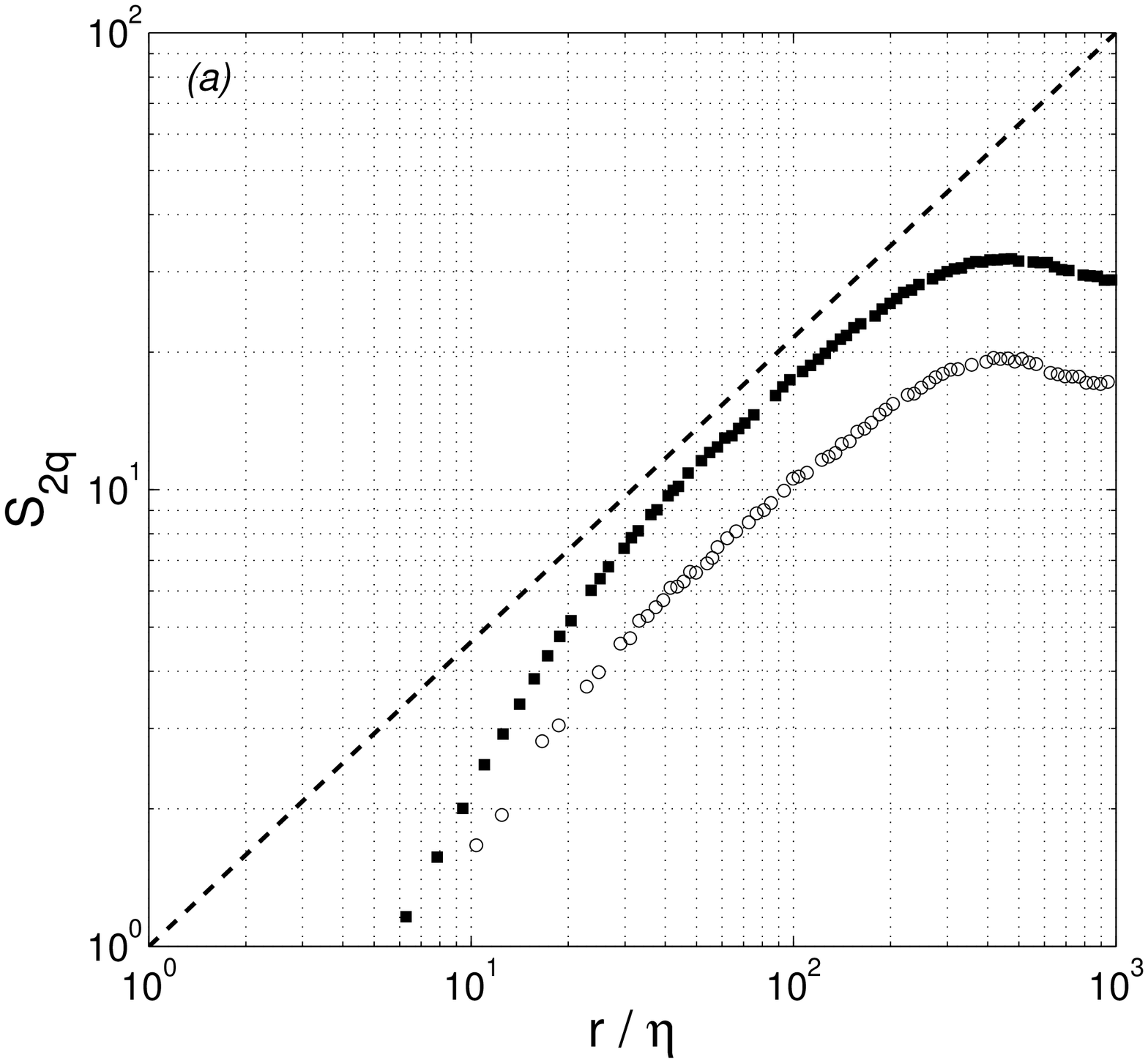}
\label{fig:s2q}}
 \subfigure
{\includegraphics[width=7.5cm]{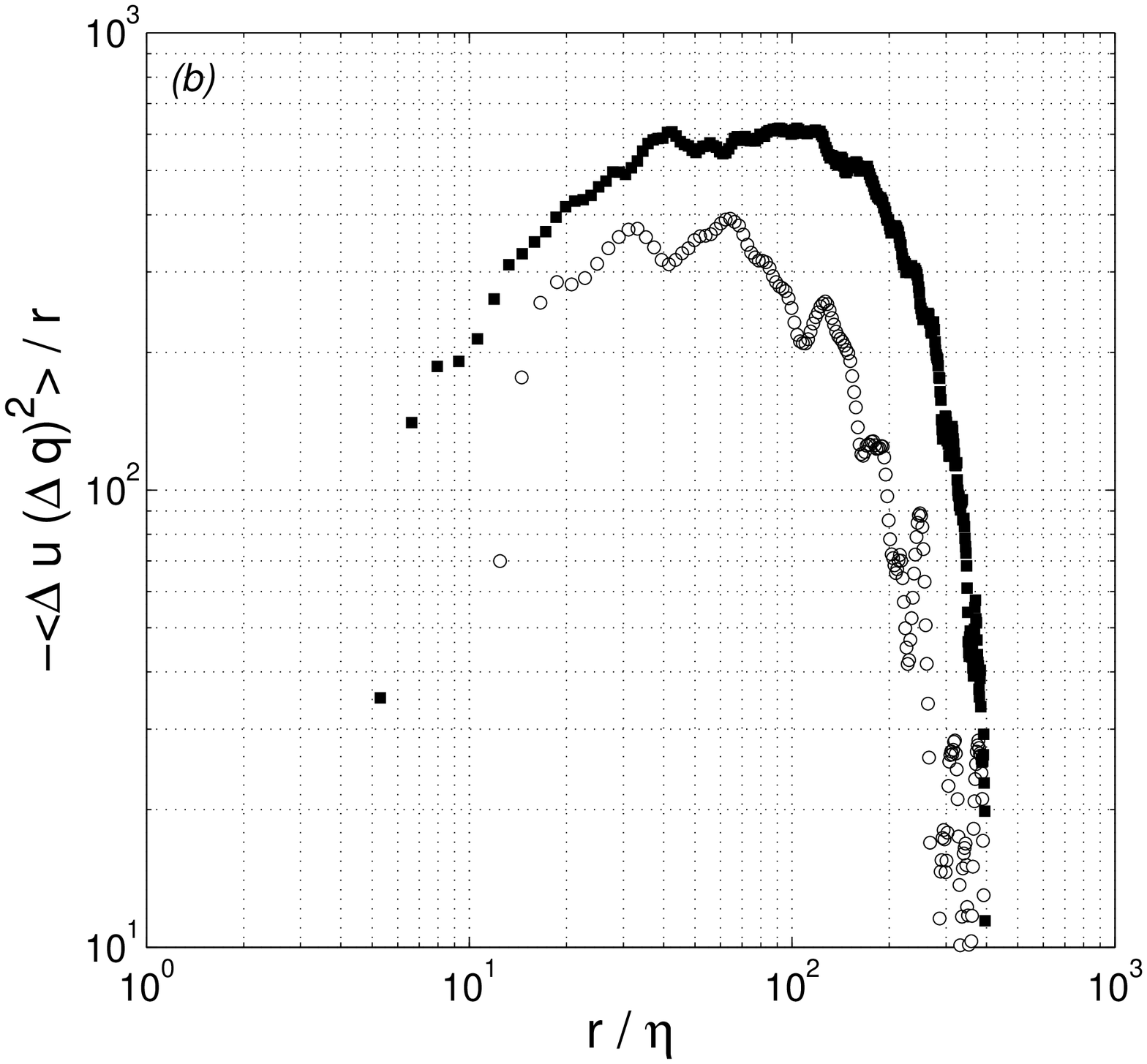}
\label{fig:s3q}} \caption{\textit{(a)} Second-order structure
functions $\langle (\Delta q)^2\rangle$. The black dotted line represents
the $r^{2/3}$ scaling. \textit{(b)} Third-order structure functions
divided by the separation $r$, $-\langle \Delta u (\Delta
q)^2\rangle/r$. Symbols: $\bigcirc$, \textit{MoSTI} injector;
$\blacksquare$, \textit{MuSTI} injector.}
\end{figure}

In the context of the present work, it is of particular relevance to
compare $S_{2q}$ for both devices, see Figure \ref{fig:s2q}, at the
same spatial location (here $x/M_3 \approx 2$). The representation is done as
a function of $r/\eta$, where $\eta$ is the Kolmogorov microscale.

The first thing to be noted is that the MuSTI device presents
(approximately twice) higher values of $S_{2q}$ than those of MoSTI,
and this holds for all the range of scales. The large-scale limit
($r\rightarrow \infty$) of $S_{2q}$ is (under the hypothesis of
homogeneity) twice the total kinetic energy $q^2$. The ratio of the
large scale limits ($\approx 1.8$) fully corresponds to the ratio of
the total kinetic energies of both devices. For both flows,
noteworthy are the energy accumulations at large scales,
approximately peaking at the scale $400 \eta$, corresponding to the
injector(s) mesh $M_3$. A further analysis of $S_{2q}$ reveals that both
of them exhibit a restricted scaling range (RSR), for which the
scaling is very nearly equal to the asymptotic value of $2/3$
(equivalent to the $-5/3$ exponent in the spectral space).

A deeper insight into the nature of the turbulent cascade, albeit in
the context of local isotropy, is provided by the third-order
structure function $ -\langle \Delta u (\Delta q)^2\rangle$, where
the velocity component $u$ is parallel to the spatial separation
$\bm{r}$, taken along the mean flow direction. The mathematical
relation between the second-and third-order structure functions,
obtained under the assumption of locally isotropic, high-Reynolds
number turbulence, is e.g. \citep{myr97},
\begin{equation}
-\langle \Delta u (\Delta q)^2\rangle +2 \nu \frac{d}{d r} \langle
(\Delta q)^2 \rangle = \frac{4}{3} \epsilon r.
\label{kolmomyr}
\end{equation}
 For very small scales, it complies with  $\epsilon_h$.
 Equation (\ref{kolmomyr}) signifies that the
energy transferred at a scale $r$ is (only) done through turbulent
advection (the third-order term) and molecular diffusion.

Figure (\ref{fig:s3q}) represents term $-\langle \Delta u (\Delta
q)^2\rangle /r$, for both devices.  This terms  should be equal to
the constant value of $4/3 \epsilon$, for high
enough Reynolds numbers. For the \textit{MoSTI} device, a very
restricted plateau is present for a range of scales lying between
$30-60$ Kolmogorov scales. On the contrary, the flow associated to
the \textit{MuSTI} device exhibits broadened turbulent energy transfer,
since term $-\langle \Delta u (\Delta q)^2\rangle /r$ is characterised by
a plateau much better defined over a range of scales $40-150$
Kolmogorov scales. This difference cannot be attributed to a Reynolds number
effect because $Re_\lambda$ is roughly the same ($\approx 220$) for
both devices at this location. This result proves that energy transfer through
turbulence is, at the same spatial position, enhanced by using
the \textit{MuSTI} injector.

As earlier emphasized, Equation (\ref{kolmomyr}) is only valid for high
Reynolds numbers, and for a limited range of scales. For moderate
Reynolds numbers associated to real flows, it is therefore of
interest to study the scale-by-scale energy transport, by also
taking into account large-scale effects (turbulent diffusion, decay,
production).  The final equation, also obtained in the context of
local homogeneity (pressure-containing terms are neglected) and
local isotropy, writes \citep{njp04}, \citep{bad05}
\begin{eqnarray}
-\langle \Delta u (\Delta q)^{2}\rangle + 2 \nu \frac{d}{dr} \langle
(\Delta q)^{2}\rangle - \frac{U}{r^{2}} \int_{0}^{r}
s^{2}  \frac{\partial}{\partial_x} \langle (\Delta q)^2 \rangle ds \nonumber \\
- \frac{1}{r^{2}} \int_{0}^{r} s^{2} \left[ \frac{1}{2} \left(
\frac{\partial}{\partial x_{\alpha}}+\frac{\partial}{\partial
x_{\alpha}^+} \left\langle (u_{\alpha} +u_{\alpha}^+)(\Delta q)^2
\right\rangle \right) \right]ds \nonumber \\
-2 \frac{\partial U}{\partial x} \frac{1}{r^{2}} \int_{0}^{r} s^{2}
\left(\langle(\Delta u)^2\rangle - \langle (\Delta v)^2\rangle
\right) ds
 =
\frac{4}{3} \epsilon r, \label{kolmogenmljet}
\end{eqnarray}
where $s$ is a dummy variable and repeated indices indicate
summation.  This equation could be written, after dividing by
 $\epsilon r$, in the following dimensionless
form

\begin{equation}
A^*+B^*+D^*+TD^*+PR^*=C^*
\end{equation}

where $C^*=4/3$, $A^*$ is the term associated to turbulent transfer,
$B^*$ to molecular effects, $D^*$ is the inhomogeneous  (`decay')
term along the streamwise direction $x$, $TD^*$ is the turbulent
diffusion and $PR^*$ is the production term.

\begin{figure}[htbp]
\centering {\includegraphics[width=10cm]{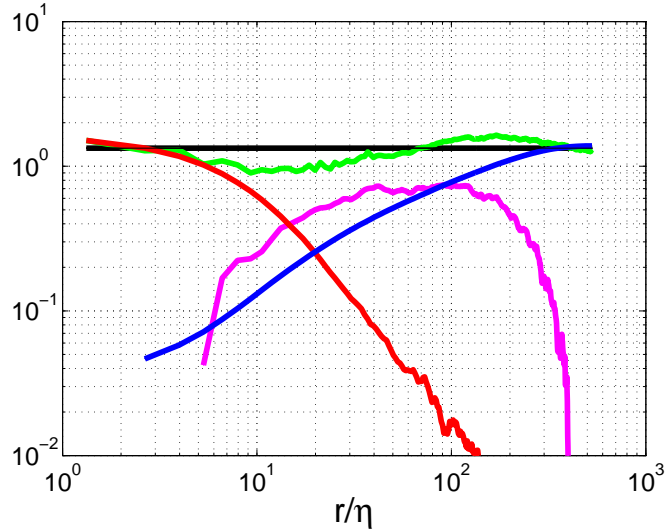}
 \caption{Scale-by-scale kinetic energy budget for the \textit{MuSTI} device.
 Term $A^*$ (magenta);  Term $B^*$ (red);  Term $D^*$ (blue);  Term $A^*+B^*+D^*$ (green).
 The black solid line represents the value of $4/3$. }
 \label{fig:budgetscale}}
\end{figure}

It is straightforward
to show that, in the context of homogeneous turbulence, the
large-scale limit of Equation (\ref{kolmogenmljet}) is Equation
(\ref{eq:budget2}). At the streamwise position investigated here,
and as previously emphasized, the effect of turbulent diffusion
compensates that of turbulent production, and therefore the 1-point
energy budget equation reduces to Equation (\ref{eq:budget3}), in which
only the decay effect has to be taken into account. In this context,
we consider that $TD^*+PR^*\approx 0$, which leads to only consider
the decay effect, as it is the case in decaying grid turbulence.
\vspace*{1ex}

Terms $A^*$, $B^*$, $D^*$ and $C^*$ are calculated for the MuSTI
injector, at the position $x/L_m=0.6$, where the total kinetic energy
is maximum, and after which the energy decay is representative of
the flow. These terms are represented in Figure \ref{fig:budgetscale},
as functions of $r/\eta$. Term $B^*$, representing energy
transferred via molecular effects, is only present at small scales,
and is negligible for large scales. Note also that for the
dissipative range domain, term $B^*$ equilibrates $C^*=4/3$, in
agreement with the definition of $\epsilon_h$. Term
$A^*$ is important for an intermediate range of scales (RSR), and
negligible for both very small and very large scales.
At this Reynolds number ($Re_{\lambda} \approx 220$), term $A^*$
does not balance $C^*$, signifying that the energy transfer is not
only performed by turbulent advection. Large-scale (here, decay)
effect represented by term $D^*$ is increasingly important for
larger and larger scales. It equilibrates term $A^*$ in the middle
of the RSR, and it is significantly larger for scales equal and
larger than the integral scale. It equilibrates by itself term $C^*$
for the largest scales of the flow, in agreement with Equation
(\ref{eq:budget2}). Finally, $A^*+B^*+D^*$ (green curve)
equilibrates term $C^*$ reasonably well over the whole range of
scales, with smaller values ($20\mbox{\%}$ lower) for the RSR scales, and
larger ($10\mbox{\%}$) for the largest scales of the RSR. This disagreement
is attributable to the effect of turbulent diffusion and production
(not considered here) and also to departures from local isotropy.
\vspace*{1ex}

We conclude here on the cold flow configuration saying that 

\begin{enumerate}
\item[(i)] the large-scales of the flow remain controlled by the largest
plate, i.e. by the spanwise interaction distance.

\item[(ii)] the small-scales are intensified in terms of energy and their
range is broadened by the multi-scale injection thanks to the enhancement
of the energy transfer term.

\item[iii)] the \textit{MuSTI} device roughly behaves as a grid
turbulence starting with $x/L_m=0.6$ position, for which all the
shear layers are well mixed, global and local isotropy hold well,
and the energy production is swept by turbulent diffusion. As far as
the \textit{MoSTI} device is concerned, the same analysis
(developed in the context of locally isotropic turbulence), does not
hold as well and therefore is not reported here.
\end{enumerate}

The issue we address in the following is the influence of the multi-scale
energy injection mode onto the interaction between combustion and intense
homogeneous and isotropic turbulence.

\subsection{Interaction with premixed combustion}

The following part is dedicated to the qualitative and quantitative
description of the interaction between the turbulence generated by
both injectors and premixed combustion. For this purpose, an air/methane
mixture is used as working fluid with an equivalence ratio $\phi = 0.6$. A
V-shaped flame is attached on a tiny heated rod located in the homogeneous
and isotropic region of the flow determined from the investigation
in non-reactive configuration. The location $x_{rod}$ of the heated
rod is given in Table \ref{tab:Turbulent} as well as the main flow
properties of the turbulent flow at the rod position.

\begin{table}[htbp]
\begin{center}
\begin{tabular}{ccccccccccc}
\hline
Injector& $\frac{x_{rod}}{L_m}$ & $H$ & $I$ & $\frac{\sqrt{q^2 / 3}}{U_c}$ (\%) & $\sqrt{q^2 / 3}$ (m/s) & $\Lambda_u (mm)$ & $\lambda (mm)$ & $\eta (mm)$ & $R_\lambda$ \\
\hline
\hline
MoSTI   & 1.45 & 0.03 & 1.18 & 14 & 0.58 & 9.1 & 1.5 & 0.10 & 64\\
MuSTI   & 1.34 & 0.04 & 1.14 & 18 & 0.88 & 6.6 & 1.3 & 0.08 & 85\\
\hline
\end{tabular}
\caption{Main properties of the turbulent flow at the location of the heated rod.}
\label{tab:Turbulent}
\end{center}
\end{table}

\begin{figure}[htbp]
\centering
\includegraphics[width=11cm]{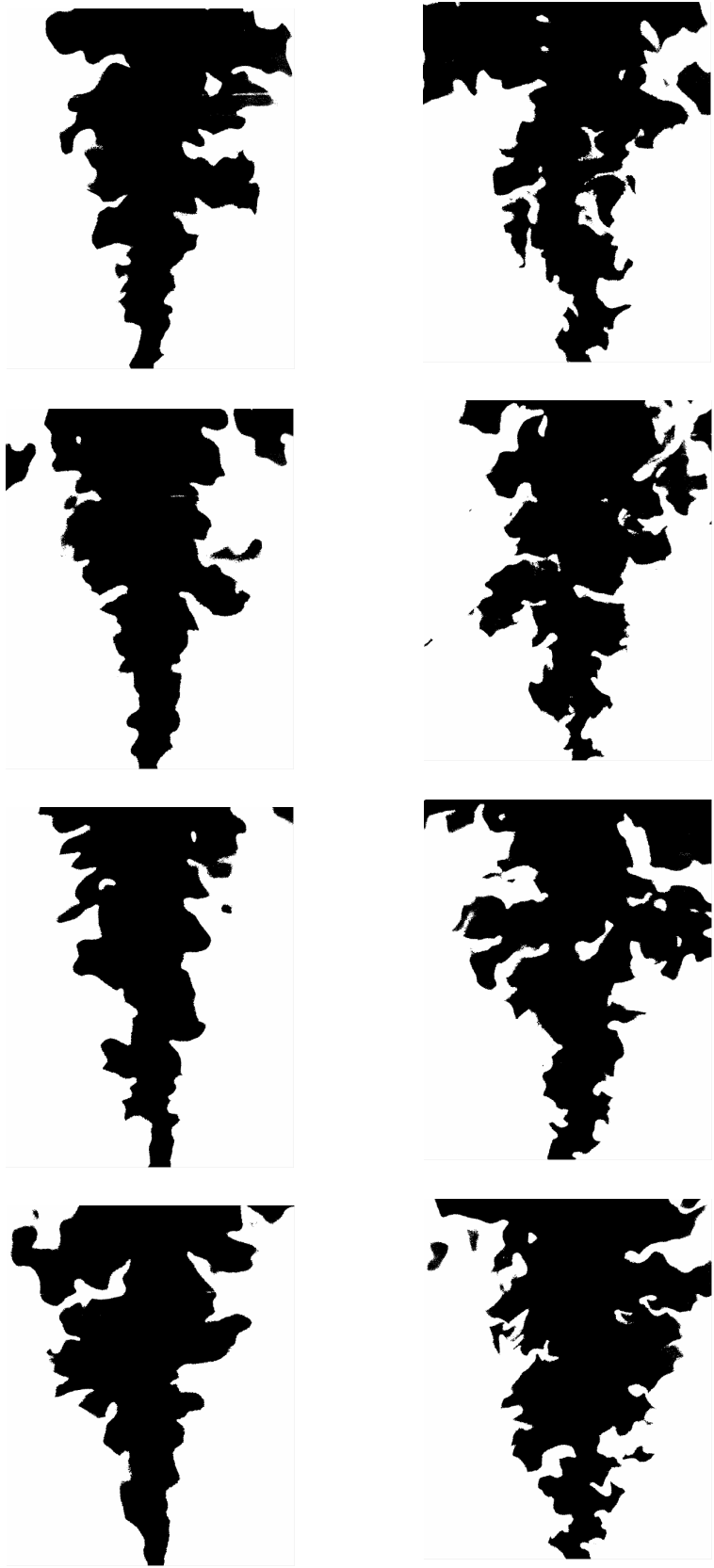}
\caption{Examples of instantaneous binary flame images recorded downstream the
\textit{MoSTI} injector (left hand-side) and the \textit{MuSTI}
injector (right hand-side).}
\label{fig:Flames}
\end{figure}

Relatively to the jet interaction origin, the heated rod is
almost located at the same position $x_{rod} / L_m$ for both
turbulence generators. At this location, the turbulent flow
generated by both injectors is nearly homogeneous and
isotropic with comparable turbulence intensity level. However,
the \textit{MuSTI} injector generates slightly smaller turbulent
scales which are more energetic than those of the \textit{MoSTI} 
device. This property improves the ability of multi-scale generated
turbulence to interact with the flame at both large- and small-scales
as evidenced by the examples of binary flame images obtained
downstream both injectors (see Figure \ref{fig:Flames}). Although
the turbulence statistics are comparable for both configurations,
one can notice an evident difference in the global shape of
the flames. The wrinkling of the
front flame edges is clearly much larger for the \textit{MuSTI}
injector as testified by the turbulent flame brush
and the local extinctions allowing unburnt gases
pockets to penetrate deeply inside the burnt gases. Such phenomena
are not visible for the \textit{MoSTI} injector.

These differences partly result from the amplification of the 
turbulence intensity as reported in previous works (e.g.
\citep{Haqetal2002}, \citep{Renouetal2002}). Combining this effect with the reduction
of turbulent length-scales can also affect significantly the flame
stretch \citep{CandelPoinsot1990} by acting of both strain rate and
flame curvature.

The competition between the chemical reaction rate and the
turbulence is characterised by:

\begin{enumerate}
\item the ratio between the integral time-scale $\tau_t =
\Lambda_u / \sqrt{q^2/3}$ and the chemical
time-scale $\tau_c$ for the large-scale turbulence.
\item the ratio between the chemical time-scale $\tau_c$ and the
Kolmogorov time $\tau_\eta = \eta / u_\eta$ (with $u_\eta =
(\nu \epsilon)^{1/4}$ the Kolmogorov velocity) for the small-scale turbulence .
\end{enumerate}

The chemical time-scale $\tau_c$ is defined as the ratio between
the laminar flame thickness $\delta_L^0$ (estimated from the maximal
temperature gradient $\delta_L^0 = \frac{T_b - T_u}{(\nabla T)_{max}}$
with $T_u$ the temperature in fresh gases and $T_b$ the temperature of
burnt gases) and the laminar flame speed $S_L^0$ \citep{PoinsotVeynante2001}.
These parameters are reported in Table \ref{tab:Chemical} for our methane/air
mixture with an equivalence ratio $\phi = 0.6$. The large-scale dynamic ratio
can be expressed via Damk{\"o}hler number

\begin{equation}
Da = \frac{\Lambda_u}{\delta_L^0} \frac{S_L^0}{\sqrt{q^2/3}}
\end{equation}

while the small-scale dynamic ratio is represented by the Karlovitz number

\begin{equation}
Ka = \left(\frac{\sqrt{q^2/3}}{S_L^0}\right)^{3/2} \left(\frac{\Lambda_u}{\delta_L^0}\right)^{-1/2}
\end{equation}

These numbers are given in Table \ref{tab:Chemical} for both injectors.

\begin{table}[htbp]
\begin{center}
\begin{tabular}{cccccc}
\hline
Injector& $\phi$ & $S_L^0$ (m/s)$^a$ & $\delta_L^0$ (mm)$^b$ & $Ka$ & $Da$ \\
\hline
\hline
MoSTI   & 0.6 & 0.11 & 1.02 & 4.09 & 1.68 \\
MuSTI   & 0.6 & 0.11 & 1.02 & 8.82 & 0.81 \\
\hline
\end{tabular}
\caption{Main chemical properties of the fuel-air mixture. $^a$ estimated from \citep{Bradleyetal1996}.
$^b$ estimated from \citep{Lafayetal2008}.}
\label{tab:Chemical}
\end{center}
\end{table}

The \textit{MuSTI} device shortens both large- and small-scale turbulent
times compared to the chemical time as testified by the decrease of Damk{\"o}hler
number $Da$ and the increase of Karlovitz number $Ka$ (by a factor $\approx 2$).

\begin{figure}[htbp]
\centering
\subfigure
%{\includegraphics[width=7.5cm]{ProgressMono.eps}
{\includegraphics[width=7cm]{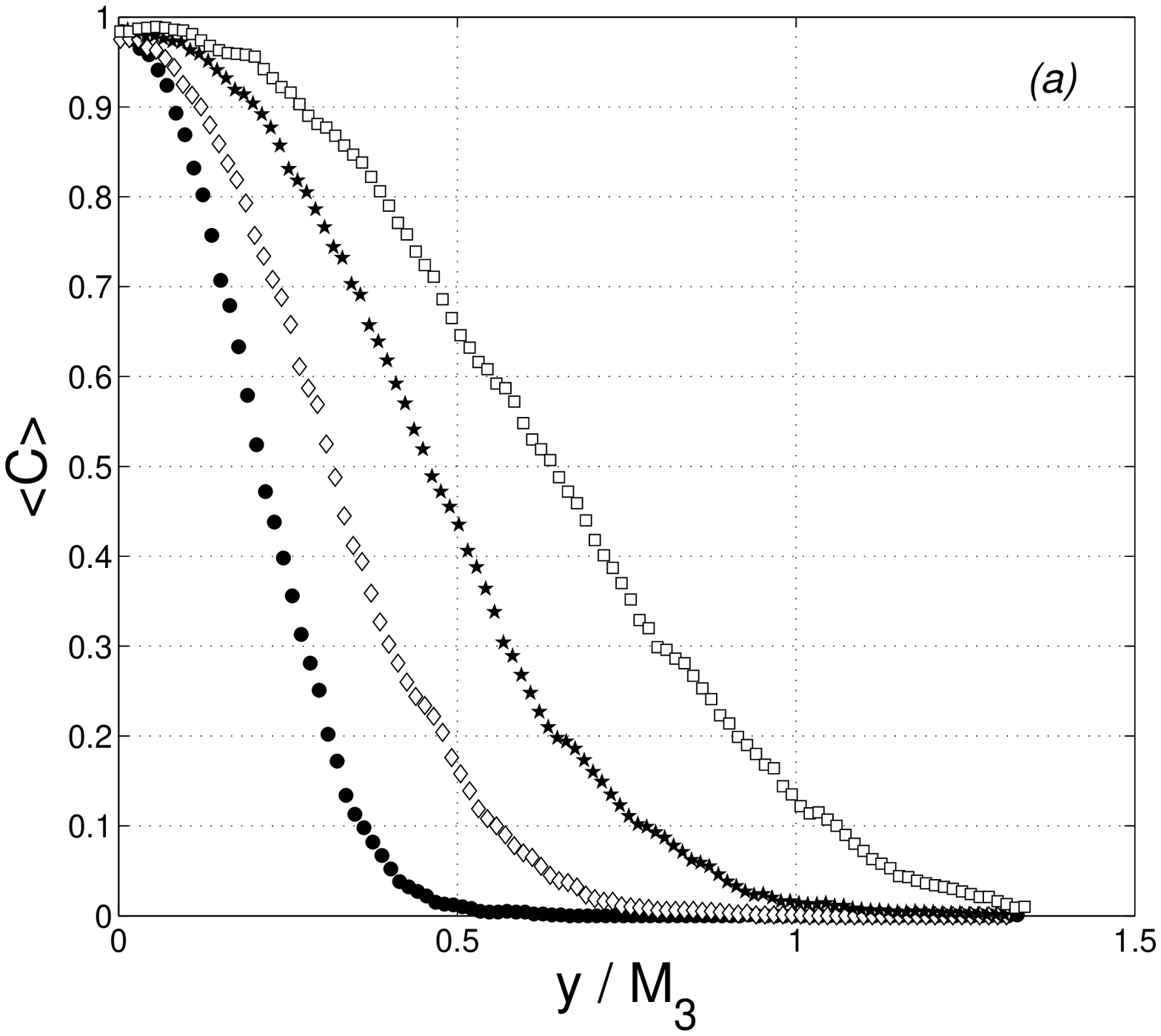}
\label{fig:ProgressMono}}  \subfigure
%{\includegraphics[width=7.5cm]{ProgressMulti.eps}
{\includegraphics[width=7cm]{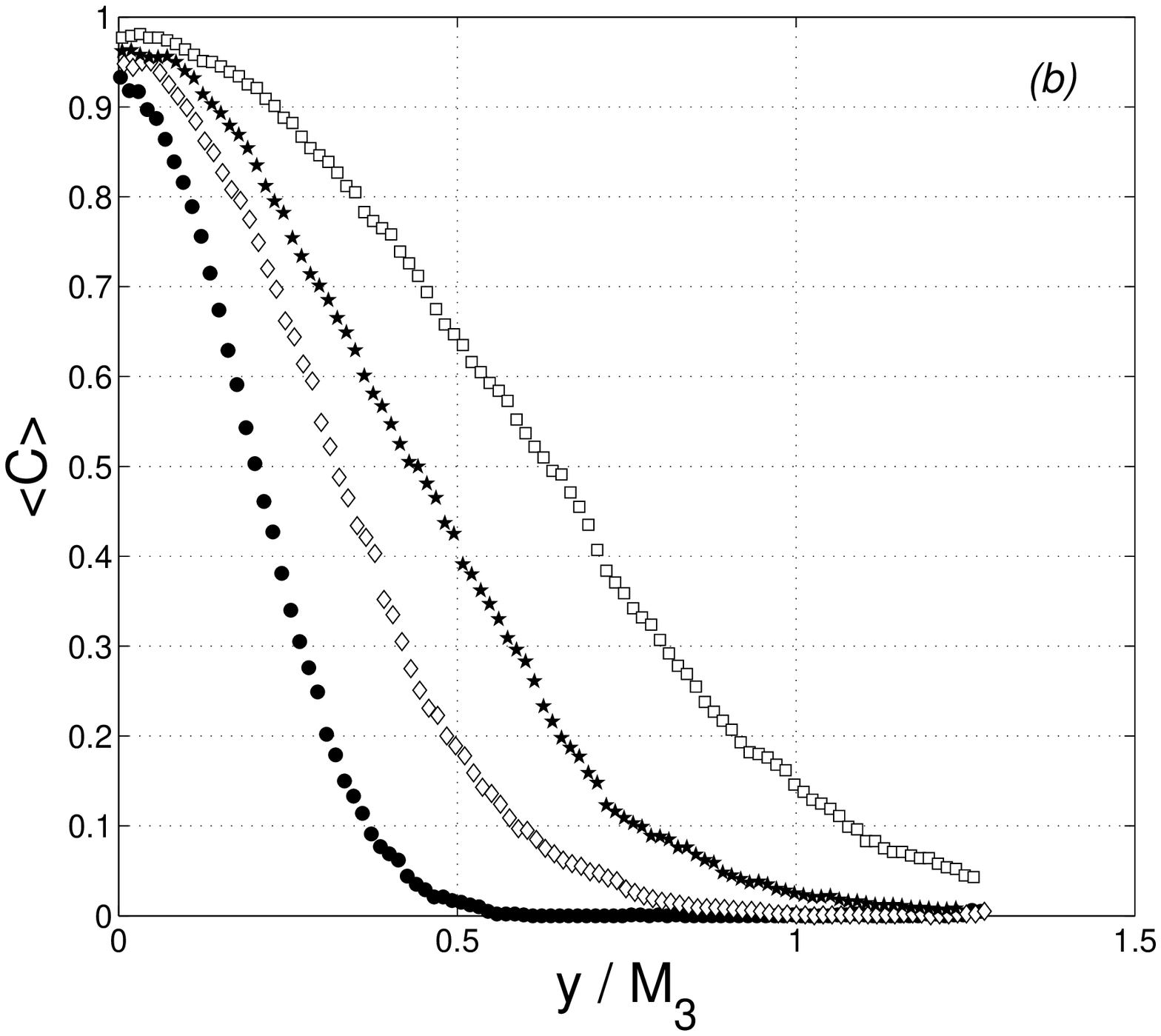}
\label{fig:ProgressMulti}} \caption{Transverse profiles of the mean
progress variable $\left\langle C\right\rangle$ at various distances
from the heated rod downstream the \textit{MoSTI} injector
\textit{(a)} and the \textit{MuSTI} injector \textit{(b)}. Symbols:
$\bullet$, $(x - x_{rod})/M_3 = 1$; $\diamond$, $(x - x_{rod})/M_3 =
1.5$; $\star$, $(x - x_{rod})/M_3 = 2$; $\square$, $(x -
x_{rod})/M_3 = 2.5$;}
\end{figure}

One of the main consequences of the turbulent dynamics amplification by
multi-scale injection mode is to increase the flame wrinkling as
illustrated in Figure \ref{fig:Flames}. This phenomenon can be
quantified by the flame brush thickness $\delta_t$ which is evaluated
from the profile of the Reynolds average progress variable
$\left\langle C\right\rangle$. This mean
progress variable distribution corresponds to a map of flame front
presence probability. It is obtained from instantaneous binarized
images ($C = 0$ in fresh gases and $ C = 1$ in burnt gases),
representing the instantaneous map of the progress variable. It
is important to notice that the spatial resolution of the tomography
technique is not good enough to investigate the local flame structure.
We restrict therefore the discussion on the global properties of
the flame, i.e. at large-scale. In this way, we implicitly assume that
the flame front is thin with respect to the spatial resolution of the measurements.
\vspace*{1ex}

In Figures \ref{fig:ProgressMono} and \ref{fig:ProgressMulti}, we report the
profiles of the mean progress variable $\left\langle C\right\rangle$
measured at several distances from the heated rod for the
\textit{MoSTI} and the \textit{MuSTI} injectors respectively. These
figures evidence the flame expansion with the distance from the
heated rod. For the \textit{MuSTI} injector, one can remark that the
maximum of the mean progress variable $\left\langle
C\right\rangle_{max}$ remains lower than unity and tends towards $1$
with increasing streamwise distance. This behaviour cannot be
attributed to measurements uncertainties but is related to the strong
wrinkling induced by the multi-scale forcing.

\begin{figure}[htbp]
\centering
\subfigure
%{\includegraphics[width=7.5cm]{Brush.eps}
{\includegraphics[width=7cm]{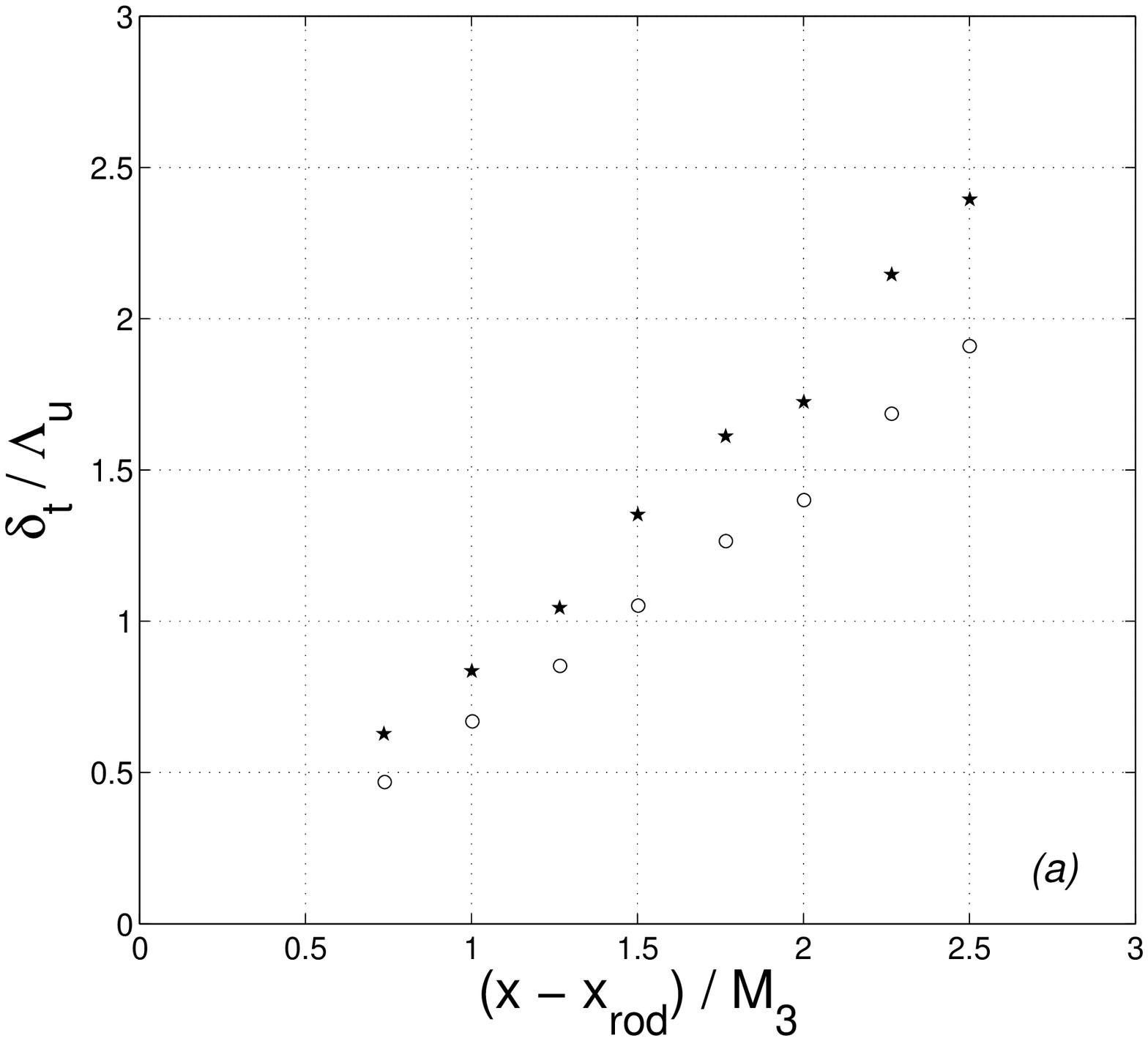} \label{fig:Brush}}
 \subfigure
%{\includegraphics[width=7.5cm]{BrushDiffusion.eps}
{\includegraphics[width=7cm]{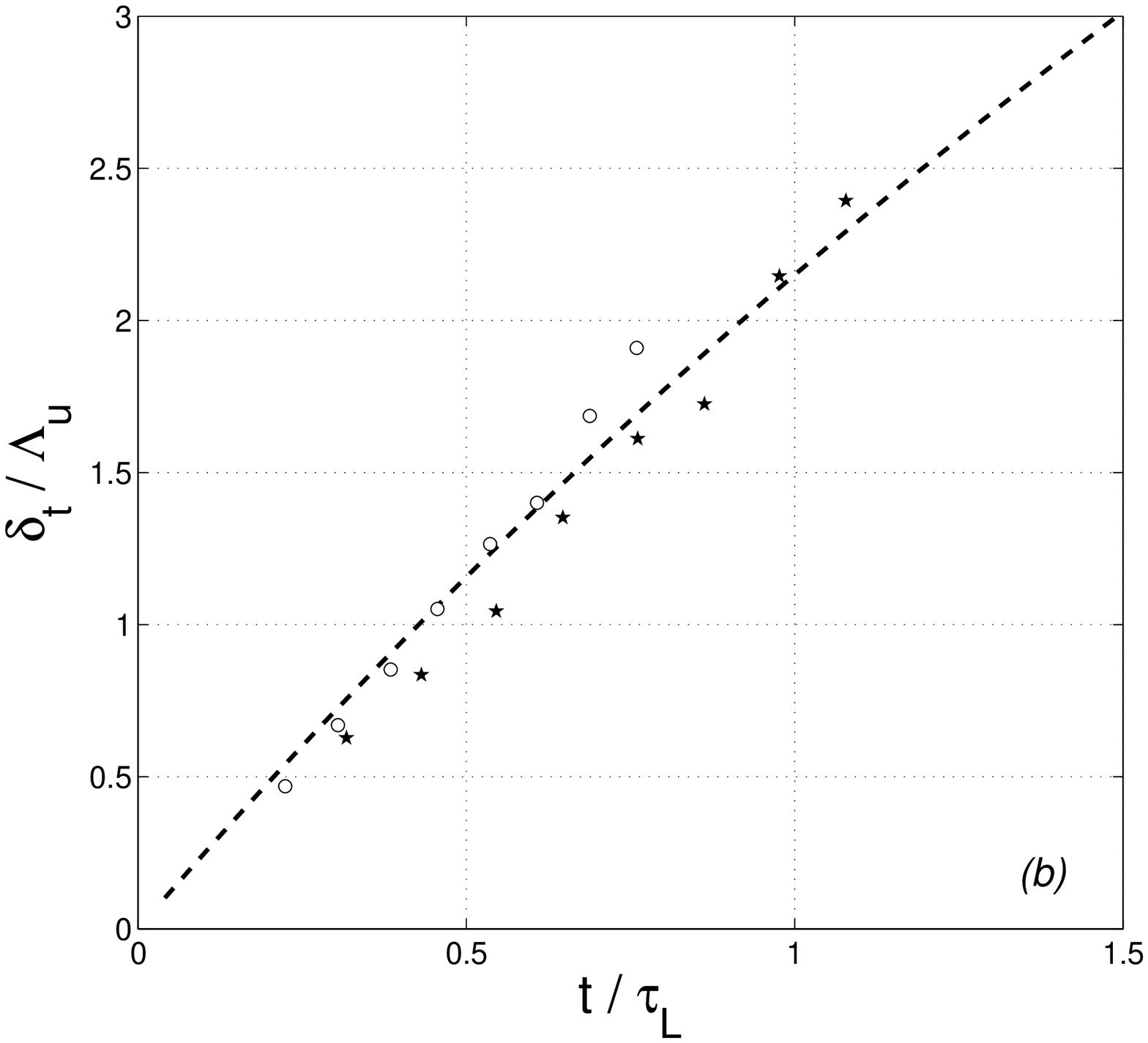}
\label{fig:BrushDiffusion}} \caption{\textit{(a)} Streamwise
variation of the normalized flame brush thickness $\delta_t /
\Lambda_u$ vs. the dimensionless distance from the heated rod
$\frac{x - x_{rod}}{M_3}$. \textit{(b)} Comparison with the Taylor's
turbulent diffusion law (dashed line) given in equation
\ref{eq:diffusion}. Symbols: $\circ$, \textit{MoSTI} injector;
$\star$, \textit{MuSTI} injector.}
\end{figure}

The flame brush thickness $\delta_t$ has been computed from the mean
progress variable $\left\langle C\right\rangle$ profiles according to
\citep{Namazianetal1986}:

\begin{equation}
\frac{1}{\delta_t} = \left(\frac{\partial \left\langle C\right\rangle}
{\partial y}\right)_{max}
\end{equation}

In the case of the \textit{MuSTI} injector, the mean progress variable
profiles have been normalized by $\left\langle C\right\rangle_{max}$ as
suggested by Lipatnikov \& Chomiak \citep{LipatnikovChomiak2002}. The
flame brush thickness $\delta_t$ scaled by the integral scale $\Lambda_u$
is plotted against the dimensionless distance from the flame-holder
$\frac{x - x_{rod}}{M_3}$ in Figure \ref{fig:Brush} for both injectors.
As expected, the flame brush thickness $\delta_t$ increases versus $x$.
Moreover, the flame brush thickness $\delta_t$ is increased by about $30$\%
for the \textit{MuSTI} injector confirming the amplification of the flame
wrinkling compared to the \textit{MoSTI} device.

The growth of the flame brush thickness $\delta_t$ can be described by
the Taylor's turbulent diffusion approach (see e.g. the review article
from Lipatnikov \& Chomiak \citep{LipatnikovChomiak2002}) expressed as:

\begin{equation}
\delta_t = \sqrt{2 \pi} \Lambda_u \left\{2 \left(\frac{t}{\tau_L}\right) \left[1 - \left(\frac{\tau_L}{t}\right) \left(1 - e^{-t / \tau_L}\right) \right] \right\}^{1/2}
\label{eq:diffusion}
\end{equation}

where $t$ is the elapsed time from the source (heated rod) and $\tau_L =
\frac{\Lambda_u}{u'}$ (with $u'$ the axial root-mean square velocity)
\citep{Corrsin1963b} is the Lagrangian integral time-scale. The comparison between
the flame brush thickness $\delta_t$ evolution and the Taylor's turbulent diffusion
law is given in Figure \ref{fig:BrushDiffusion} for both injectors. Experimental
data fairly well collapse on diffusion law of Equation (\ref{eq:diffusion}) indicating
that the flame brush is mainly controlled by the large-scale turbulence surrounding
the flame. Our results are in good agreement with those reported by Goix \& al.
\citep{Goixetal1988} and Renou \& al. \citep{Renouetal2002} in grid-generated turbulence
for various flammable mixtures.
\vspace*{1ex}

\begin{figure}[htbp]
\centering
\includegraphics[width=11cm]{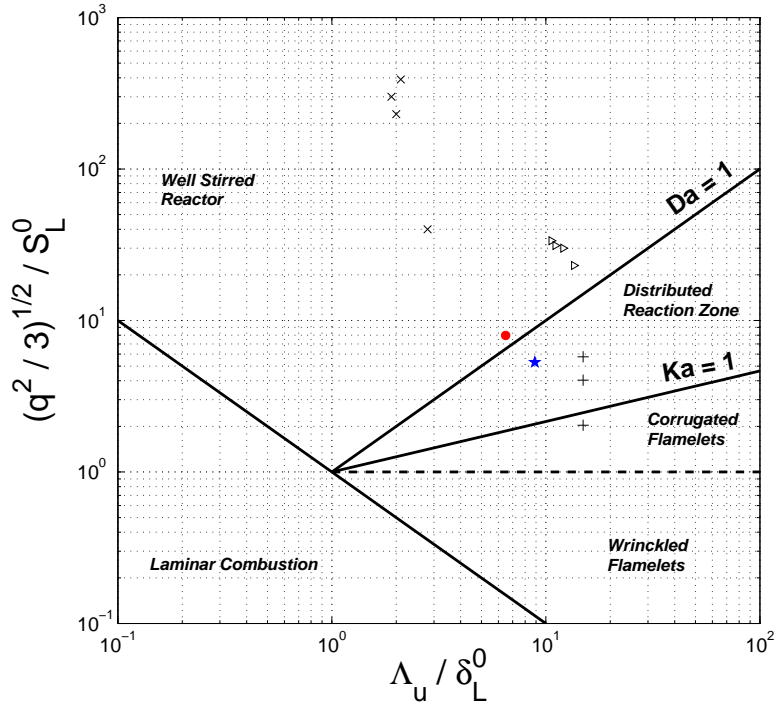}
\caption{Location of the expected flames in the combustion diagram.
Present work : ($\color{red}{\bullet}$) \textit{MuSTI} injector,
($\color{blue}{\star}$) \textit{MoSTI} injector. ($\times$) Dunn \&
al. \citep{Dunnetal2007}, ($+$) O'Young \& Bilger
\citep{OYoungBilger1997}, ($\triangleright$) Dinkelacker \& al.
\citep{Dinkelackeretal1998}.} \label{fig:CombDiag}
\end{figure}

We conclude this section by presenting the expected area covered by
the flame generated downstream the \textit{MoSTI} injector and the
\textit{MuSTI} device in Figure \ref{fig:CombDiag}. The reported
point are computed from the turbulence parameters evaluated at the
position $x_{rod}$ (see Table \ref{tab:Turbulent}) and the chemical
properties of the mixture we have used (see Table
\ref{tab:Chemical}). For comparison, experimental results obtained
by Dunn \& al. \citep{Dunnetal2007} for a piloted jet burner, O'Young
\& Bilger \citep{OYoungBilger1997} for bluff-boby and Dinkelacker \&
al. \citep{Dinkelackeretal1998} for swirl flow are also reported.
The flame generated downstream the
\textit{MoSTI} is expected to be in the distributed region, whilst
that of the \textit{MuSTI} device is expected to reach the
well-stirred reactor region. The latter is only accessible to highly
turbulent flows such as jets \citep{Dunnetal2007} or bluff-body
\citep{OYoungBilger1997} turbulence which are flows where homogeneity
and isotropy assumptions are far to be fulfilled unlike the
turbulent flow generated by the \textit{MuSTI} injector. However, a
straightforward interpretation of these results might be risky.
Indeed, combustion diagrams are mainly based on dimensional
phenomenology as well as empirical considerations. This means that
the frontiers between premixed flame regimes are not so obvious due
to the complex interactions between combustion and turbulence.
%Studying vortex/flame interactions via two-dimensional Direct
%Numerical Simulations, Poinsot \& al. \citep{Poinsotetal1990} pointed
%out some failures in the classical combustion diagrams. They have
%shown that such diagrams \textit{``underestimate the resistance of
%flame fronts to vortices, mainly because they neglect viscous,
%transient and curvature effects''}. The authors introduced the
%concept of spectral diagram to take into account the entire range of
%turbulent scales interacting with the flame. In the spectral
%diagram, the Klimov-Williams limit ($Ka = 1$) is shown to be
%appropriate to describe the transition between the flamelet regime
%and the distributed reaction zone. This claim has been confirmed by
%B{\'e}dat \& Chen \citep{BedatCheng1995} who investigated methane/air
%premixed flames at various turbulent intensities permitting a priori
%to cross the Klimov-Williams limit ($Ka = 1$). Using Rayleigh
%scattering technique, the authors studied the flame front
%properties. Their results did not revealed relevant difference
%between the flamelet regime and the distributed reaction zone.
%\vspace*{1ex}

In the frame of the present study, we cannot reveal the real impact of
the multi-scale injection onto the local flame front properties although
the flame structure seems to be largely influenced by the
\textit{MuSTI} device. One might expect that the latter, by reducing
and reinforcing the turbulent small-scales, would enable them to
compete against heat release which enhances dissipation via viscous
effects \citep{Nadaetal2004}. This process might favour the penetration of the turbulent
small-scales into the flame leading therefore to the thickening of the
pre-heated zone. In future, we aim to extend the present work on
resolved investigation of the flame front by using Rayleigh
scattering technique \citep{Lafayetal2008} in order to check the
influence of the multi-scale injection onto the flame thickness,
especially the pre-heated zone.

\section{Conclusions}

The turbulent flow produced by a new kind of turbulence generator,
and its interaction with premixed combustion, has been
experimentally investigated. This original injector is made of a
combination of three perforated plates shifted in space such that
both their hole's diameter and their blockage ratio
increase with increasing distance. This device enables a delayed
multi-scale injection of energy which has been compared with a
reference mono-scale injector (individual perforated plate).
This qualitative and quantitative comparison has been performed
in both a cold and a reactive flow.

\subsection{Main conclusions in the cold flow configuration}

Our results leads to the following observations for
the \textit{MuSTI} injector:

\begin{enumerate}
\item[(i)] the homogeneous and isotropic region is reached
'earlier', by about $50$\%, with isotropy level comparable to
standard grid-generated turbulence.
\item[(ii)] The merging length $L_m$, which corresponds to the jet
interaction origin, is found to be the relevant length-scale
characterizing the streamwise variation of the turbulent flow.
\item[(iii)] The homogeneous and isotropic region is characterised
by much larger turbulent intensity ($\approx 15$\%) than
standard grid-generated turbulence.
\item[(iv)] Due to the multi-scale injection, the turbulent kinetic energy
supply is distributed over the whole range of scales (large till the
smallest) as emphasized by second-order structure functions.
\item[(v)] In the decay region, the turbulent energy transfer at a given
scale $r$ is enhanced, as underlined by the third-order structure
function $-\langle \Delta u (\Delta q)^2\rangle$.
\end{enumerate}

The scenario we propose for explaining these results is intimately
connected to the intrinsic concept of our multi-scale energy
injector. Indeed, the turbulence genesis of the multi-scale injector
is done gradually, over a much longer distance than the \textit{MoSTI}
(for which, the turbulence injection is only done along the plate
thickness). Indeed, for the \textit{MuSTI} injector, the turbulence
birth begins $45$mm ($= 3.5 M_1+ 3.5 M_2$) upstream the inlet of the
wind tunnel. Along this distance the turbulence is gradually reinforced.
In connection with the particular energy injection mode (stagnating
the flow over a large surface), fluctuations of $u$ along $x$ are
principally created in a first time. The fact that this energy supply is done
progressively in the flow, the fluctuating velocity $u$ has enough
'time/space' to feed the other two fluctuating velocity components
$v$ and $w$. Therefore, the multi-scale injector significantly
enhances transverse fluxes accelerating the spatial redistribution
of energy, and therefore reaching homogeneity and isotropy more
rapidly.

Besides the enhancement of transverse exchanges at large-scale, our
work shows that the multi-scale injection reinforces the small-scales
by accelerating and amplifying the energy transfer. This result might
be interpreted as a turbulent cascade by-pass.

\subsection{Main conclusions in the reactive flow configuration}

The interaction of premixed combustion and turbulence generated by
multi-scale and mono-scale injector has been qualitatively
investigated by front flame visualization. The multi-scale generated
turbulence strongly increases the wrinkling of the flame edge
provoking local extinctions and propagation of unburnt pockets in
the burnt gases. This observation is confirmed by the $30$\%
increase of the flame brush thickness. The flame produced
downstream the multi-scale injector is expected to reach flame
regime unaccessible to standard grid-generated turbulence in the
combustion diagram. This regime requires highly turbulent level
which are usually incompatible with homogeneity and isotropy. The
multi-scale injector overcomes these drawbacks by fulfilling the
aforementioned conditions. However, due to experimental
restrictions, our analysis of flame properties cannot reveal the
real influence of the multi-scale injection onto the local flame front.
Future works will be specifically dedicated to the investigation of
the local flame front, in particular the interaction between the
multi-scale generated turbulence and the pre-heated zone.

\vskip 1truecm

\noindent
{\bf Acknowledgments:} The authors are grateful to Dr. G. Godard for his very useful
help with the LDV measurement system, to Dr. F. Corbin for his
assistance with the LDV acquisition system. This work has been
supported by a R{\'e}gion Haute-Normandie grant.

%%%%%%%%%%%%%%%%%%%%%%%%%%%%%%%%%%%%%%%%%%%%%%%%%%%%%%%%%%%%%%%%%%%%%%%%%%%%%%%%%%%%%%%%%%%%%%%%%%%%%%%%%%%%%%%%%%%%%%%%%%%
%%%%%%%%%%%%%%%%%%%%%%%%%%%%%%%%%%%%%%%%%%%%%%%%%%%%%%%%%%%%%%%%%%%%%%%%%%%%%%%%%%%%%%%%%%%%%%%%%%%%%%%%%%%%%%%%%%%%%%%%%%%
%%%%%%%%%%%%%%%%%%%%%%%%%%%%%%%%%%%%%%%%%%%%%%%%%%%%%%%%%%%%%%%%%%%%%%%%%%%%%%%%%%%%%%%%%%%%%%%%%%%%%%%%%%%%%%%%%%%%%%%%%%%
%\newpage
%\singlespacing
\begin{small}
\bibliography{}

\begin{thebibliography}{99}
    
    \bibitem{Pitsch2006}
    H. Pitsch, {\em Large Eddy Simulation of turbulent combustion},
    Ann. Rev. Fluid Mech. 38 (2006), pp. 453--482
    
    \bibitem{Peters1986}
    N. Peters,
    {\em Laminar flamelets concepts in turbulent combustion},
    21st Symp. on Combustion (1986), pp. 1231--1250
    
    \bibitem{BorghiChampion2000}
    R. Borghi and M. Champion,
    {\em Mod{\'e}lisation et th{\'e}orie des flammes},
    Technip, Paris (2000)
    
    \bibitem{Poinsotetal1990}
    T. Poinsot, D. Veynante and S. Candel,
    {\em Diagrams of premixed turbulent combustion based on direct simulation},
    23rd Symp. on Combustion (1990), pp. 613--619
    
    \bibitem{OYoungBilger1997}
    F. O'Young and R.W. Bilger,
    {\em Scalar gradient and related quantities in turbulent premixed flames},
    Comb. Flame 109 (1997), pp. 682--700
    
    \bibitem{Dinkelackeretal1998}
    F. Dinkelacker, A. Soika, D. Most, D. Hofmann, A. Leipertz, W. Polifke, K. D{\"o}bbeling,
    {\em Structure of locally quenched highly turbulent lean premixed flame},
    Proc. Combust. Inst. 27 (1998), pp. 857--865

    \bibitem{Dunnetal2007}
    M.J. Dunn, A.R. Masri and R.W. Bilger,
    {\em A new piloted premixed jet to study strong finite-rate chemistry effects},
    Comb. Flame 151 (2007), pp. 46--60
    
    \bibitem{Vervischetal2004}
		L. Vervisch, R. Hauguel, P. Domingo and M. Rullaud,
		\textit{Three facets of turbulent combustion modelling: DNS of premixed V-flame,
		LES of lifted nonpremixed flame and RANS of jet-flame},
		J. Turbul. 5 (2004)
		
		\bibitem{Nadaetal2004}
    Y. Nada, M. Tanahashi and T. Miyauchi,
    \textit{Effect of turbulence characteristics on local flame structure of
    H$_2$-air premixed flames},
    J. Turbul. 5 (2004)
    
    \bibitem{BatchelorTownsend1948}
    G.K. Batchelor and A.A. Townsend,
    \textit{Decay of isotropic turbulence in the initial period},
    Proc. R. Soc. Lond. A 193 (1948), pp. 539--558
    
    \bibitem{Corrsin1963a}
    S. Corrsin,
    \textit{Turbulence: experimental methods. In Handbuch der Physik},
    Springer (1963), pp. 524--589
    
    \bibitem{ComtebellotCorrsin1966}
    G. Comte-Bellot and S. Corrsin,
    \textit{The use of a contraction to improve isotropy of grid-generated turbulence},
    J. Fluid Mech. (1966) 25, pp. 657--682
    
    \bibitem{GadelHakCorrsion1974}
    M. Gad-el-Hak and S. Corrsin,
    \textit{Measurements of the nearly isotropic turbulence behind a uniform jet grid},
    J. Fluid Mech. 62 (1974), pp. 115--143
    
    \bibitem{Makita1991}
    H. Makita,
    \textit{Realization of a large-scale turbulence field in a small wind tunnel},
    Fluid Dyn. Res. 8 (1991), pp. 53--64
    
    \bibitem{MW1996}
    L. Mydlarski and Z. Warhaft,
    \textit{On the onset of high-Reynolds-number grid-generated wind tunnel turbulence},
    J. Fluid Mech. 320 (1996), pp. 331-368
    
    \bibitem{HurstVassilicos2007}
    D. Hurst and J.C. Vassilicos,
    \textit{Scalings and decay of fractal-generated turbulence},
    Phys. Fluids 19 (2007), 035103
    
    \bibitem{TanAtichatetal1982}
    J. Tan-Atichat, H.M. Nagib and R.I. Loehrke,
    \textit{Interaction of free-stream turbulence with screens and grids: a balance
    between turbulent scales},
    J. Fluid Mech. 114 (1982), pp. 501--528
    
    \bibitem{GrothJohansson1988}
    J. Groth and A.V. Johansson,
    \textit{Turbulence reduction by screens},
    J. Fluid Mech. 197 (1988), pp. 139--155
    
    \bibitem{Degardinetal2006}
    O. D{\'e}gardin, B. Renou and A.M. Boukhalfa,
    \textit{Simultaneous measurement of temperature and fuel mole fraction using acetone
    planar induced fluorescence and Rayleigh scattering in stratified flames},
    Exp. Fluids 40 (2006), pp. 452-463
    
    \bibitem{Foucaultetal2004}
    J.M. Foucault, J. Carlier and M. Stanislas,
    \textit{PIV optimization for the study of turbulent flow using spectral analysis},
    Meas. Sci. Tech. 15 (2004), pp. 1046--1058
    
    
%    \bibitem{Rossietal2006}
%    L. Rossi, J.C. Vassilicos and Y. Hardalupas,
%    \textit{Electromagnetically controlled multiple scale flow},
%    J. Fluid Mech. 558 (2006), pp. 207--242
        
    \bibitem{MazziVassilicos2004}
    B. Mazzi and J.C. Vassilicos,
    \textit{Fractal-generated turbulence},
    J. Fluid Mech. 502 (2004), pp. 65--87
    
    \bibitem{VillermauxHopfinger1994}
    E. Villermaux and E.J. Hopfinger,
    \textit{Periodically arranged co-flowing jets},
    J. Fluid Mech. 263 (1994), pp. 63--92
    
    \bibitem{LawsLivesey1978}
    E.M. Laws and J.L. Livesey,
    \textit{Flow through screens},
    Ann. Rev. Fluid Mech. 10 (1978), pp. 247--266
    
    \bibitem{TennekesLumley1972}
    H. Tennekes and J.L. Lumley,
    \textit{A first course in turbulence},
    MIT Press, 1972    

		\bibitem{Pope2000}
    S.B. Pope,
    \textit{Turbulent flows},
    Cambridge University Press, 2000

		\bibitem{Lavoieetal2007b}
    P. Lavoie, G. Avallone, F. De Gregorio, G.P. Romano and R.A. Antonia,
    \textit{Spatial resolution of PIV for the measurement of turbulence},
    Exp. Fluids 43 (2007), pp. 39--51

		\bibitem{Danailaetal2002}
    L. Danaila, F. Anselmet and R.A. Antonia,
    \textit{An overview of the effect of large-scale nonhomogeneities on
    small-scale turbulence},
    Phys. Fluids 14 (2002), pp. 2475--2484
        
		\bibitem{myr97}
    R.A. Antonia, M. Ould-Rouis,  F. Anselmet and Y. Zhu,
    \textit{Analogy between predictions of Kolmogorov and Yaglom},
    J. Fluid Mech. 332 (1997), pp. 395--409

		\bibitem{njp04}
    L. Danaila, R.A. Antonia  and P. Burattini,
    \textit{Progress in studying small-scale turbulence using exact
    two-point equations},
    New J. Phys 128 (2004), pp. 1--23

		\bibitem{bad05}
    P. Burattini, R.A. Antonia and L. Danaila,
    \textit{Similarity in the far field of a turbulent round jet},
    Phys. Fluids 17 (2005), 025101

		\bibitem{Haqetal2002}
    M.Z. Haq, C.G.W. Sheppard, R. Woolley, D.A. Greenhalgh and R.D. Lockett,
    \textit{Wrinkling and curvature of laminar and turbulent premixed flames},
    Comb. Flame 131 (2002), pp. 1--15

		\bibitem{Renouetal2002}
    B. Renou, A. Mura, E. Samson and A. Boukhalfa,
    \textit{Characterization of the local flame structure and the flame surface
    density in freely propagating premixed flames at various Lewis numbers},
    Combust. Sci. Tech. 174 (2002), pp. 143--179

		\bibitem{CandelPoinsot1990}
    S. Candel and T. Poinsot,
    \textit{Flame stretch and the balance equation for the flame surface area},
    Comb. Sci. Tech. 70 (1990), pp. 1--15
    
		\bibitem{PoinsotVeynante2001}
    T. Poinsot, D. Veynante,
    Theoritical and Numerical Combustion (2001) Edwards.
    
    \bibitem{Bradleyetal1996}
    D. Bradley, P.H. Gaskell, X.J. Gu,
    {\em Burning velocities, markstein lengths and flame quenching for spherical
    methane-air flames: a computational study}
    Combust. Flame 104 (1996) 176-198.
    
    \bibitem{Lafayetal2008}
    Y.Lafay, B. Renou, G. Cabot and M. Boukhalfa,
    \textit{Experimental and numerical investigation of the effect of H2 enrichment
    on laminar methane-air flame thickness},
    Comb. Flame 153 (2008), pp. 540--561
        
    \bibitem{Namazianetal1986}
    M. Namazian, I.G. Shepherd and L. Talbot,
    \textit{Characterization of the density fluctuations in turbulent V-shaped
    premixed flames},
    Combust. Flame 64 (1986), 299
    
    \bibitem{LipatnikovChomiak2002}
    A.N. Lipatnikov and J. Chomiak,
    \textit{Turbulent flame speed and thickness: phenomenology, evaluation and
    application in multi-dimensional simulations},
    Prog. Energy Comb. Sci. 28 (2002), pp. 1--74
    
    \bibitem{Corrsin1963b}
    S. Corrsin,
    \textit{Estimates of the relation between Eulerian and Lagrangian scales in
    large Reynolds number turbulence},
    J. Atmosph. Sci. 20 (1963), pp. 115--119
    
    \bibitem{Goixetal1988}
    P. Goix, P. Parantho{\"e}n and M. Trinit{\'e},
    \textit{A tomographic study of measurements in a V-shaped $\mbox{H}_2$-air
    flame and a Lagrangian interpretation of the turbulent flame brush evolution},
    Comb. Flame 81 (1988), pp. 229--241    

%    \bibitem{BedatCheng1995}
%    B. B{\'e}dat and R.K. Cheng,
%    \textit{Experimental study of premixed flames in intense isotropic turbulence},
%    Comb. Flame 100 (1995), pp. 485--494
\end{thebibliography}
    
\end{small}
%\label{lastpage}

\end{document}